\documentclass[intlimits,twoside,a4paper]{article}

\usepackage{amsmath,amssymb}
\usepackage{graphicx}
\usepackage{wrapfig}
\usepackage{color}

\usepackage[T2A]{fontenc}
\usepackage[cp1251]{inputenc}

\usepackage[eqsecnum]{cmpj2}
\newcommand{\Tr}{\mathop{\rm Tr}\nolimits}
\newcommand{\Adj}{\mathop{\rm Adj}\nolimits}
\newcommand{\diag}{\mathop{\rm diag}\nolimits}

\issue{2015}{18}{2}{23802}
\doinumber{10.5488/CMP.18.23802}

\title[Adsorption of symmetric random copolymer onto symmetric random surface]%
{Adsorption of symmetric random copolymer onto symmetric random surface:
the annealed case}
\author[A.A. Polotsky]{A.A. Polotsky\refaddr{label1,label2}}
\addresses{
\addr{label1} Institute of Macromolecular Compounds, Russian Academy of Sciences,
  31 Bolshoy prospekt, 199004 Saint Petersburg, Russia
\addr{label2} Saint Petersburg National Research University of Information Technologies,
  Mechanics and Optics (ITMO University), 49 Kronverkskiy prospekt,
  197101 Saint Petersburg, Russia
}
\date{Received September, 24, 2014, in final form February 4, 2015}

\begin{document}

\maketitle

\begin{abstract}
Adsorption of a symmetric (\emph{AB}) random copolymer (RC) onto a
symmetric (\emph{ab}) random heterogeneous surface (RS) is studied
in the annealed approximation by using a two-dimensional partially
directed walk model of the polymer. We show that in the symmetric
case, the expected \emph{a posteriori} compositions of the RC and the
RS have correct values (corresponding to their\emph{ a priori} probabilities)
and do not change with the temperature, whereas second moments of
monomers and sites distributions in the RC and RS change. This indicates
that monomers and sites do not interconvert but only rearrange in
order to provide better matching between them and, as a result, a
stronger adsorption of the RC on the RS. However, any violation of
the system symmetry shifts equilibrium towards the major component
and/or more favorable contacts and leads to  interconversion of monomers and sites.
\keywords random copolymer, random surface, polymer adsorption, annealed approximation,
generating functions
\pacs 87.10.Hk, 82.35.Gh, 82.35.Jk
\end{abstract}

\section{Introduction}

Adsorption of heteropolymers~--- polymers composed of monomer units of
two and more types~--- onto chemically heterogeneous surfaces was intensively
studied in the last decades. A particular interest to this problem
is motivated by its connection to the question of molecular recognition
playing a crucial role in living organisms and in various biomedical/biotechnological
applications. To understand the mechanisms of the polymer-surface
recognition, the problem was extensively investigated from different
angles by using relatively simple and physically transparent models.

In particular, directed walk models of polymers \cite{Privman:1989,Janse_van_Rensburg:2000,Janse_van_Rensburg:2003}
played an important role in studying homopolymer adsorption \cite{Privman:1988-2,Carvalho:1988,Forgacs:1989,Brak:2005,Brak:2007,Owczarek:2008,Owczarek:2009,Brak:2010,Iliev:2011-2}
and collapse \cite{Brak:1992,Owczarek:1993,Zhou:2006,Owczarek:2007,Nguyen:2013}
and related problems of force-induced desorption \cite{Orlandini:2004-1,Orlandini:2004-2,Orlandini:2009,Iliev:2010,Orlandini:2010,Osborn:2010,Owczarek:2010,Tabbara:2012,Iliev:2012-1}
and unfolding of a collapsed macromolecule \cite{Brak:2009,Lam:2010}.
Random \cite{Orlandini:2002,Soteros:2004,Polotsky:2012} and periodic
\cite{Whittington:1998,Iliev:2011-1,Polotsky:2014} copolymer adsorption
and mechanical desorption \cite{Iliev:2004,Iliev:2012-2,Iliev:2013}
were also studied with the aid of directed models. The advantage of
directed models consists in their simplicity; polymer directedness
allows one to obtain an exact solution in most cases, typically in the
long chain limit. At the same time, they provide a physically reasonable
and tractable picture of the phenomenon under study which is commonly
in agreement with the results of a more realistic computer simulation of
the same (or similar) system in 3 dimensions. Another attractive feature
of directed polymer models is the inherent self-avoidance of polymer
conformations (and the corresponding impossibility of visiting the
same surface site by two different monomer units simultaneously).

To study homo- or heteropolymer adsorption onto a homogeneous
surface, one can use fully directed polymer models (in particular,
Dyck or Motzkin paths). In the case when both the polymer and the
surface are heterogeneous and the polymer should adjust its conformation
to the surface pattern in order to attain a better matching of its monomer
sequence to the heterogeneous surface pattern and maximize the amount of favorable polymer-surface
contacts, the minimal model that allows one to consider this phenomenon
is the two-dimensional partially directed walk (2D-PDW) model. The
fully directed model, which is simpler than the 2D-PDW one, is not
suitable for this purpose because for a particular monomer unit, there is
one and only one surface site that it can visit (hence, heteropolymer
adsorption onto heterogeneous surface is equivalent to the situation
of a heteropolymer adsorption onto a homogeneous surface).

In our recent paper \cite{Polotsky:2012}, a random copolymer (RC)
adsorption onto a random surface (RS) was considered in the framework
of the  2D-PDW model of the polymer on a square lattice (the ``surface'',
therefore, was simply a line).  In order to take  correlations into account,
both random sequences of monomers (in the RC) and sites (in the RS)
were modelled as first-order Markov chains. The problem was solved
by using a combination of the annealed approximation to perform
double averaging over sequence and surface disorder, and the generating
functions (GFs) approach to sum over all conformations of the RC.
The key result of the work \cite{Polotsky:2012} was the derivation
of  an equation to find the smallest singularity of the GF
of the adsorbed chain. The latter provides an asymptotic form of
the canonical partition function for the annealed system which, in
turn, gives an access to the calculation of various observables. This
equation was then applied in \cite{Polotsky:2012} for the analysis
of the adsorption transition point for different sets of the system
parameters. This allowed us to study the effect of the interplay between
correlations in the RC and the RS on the transition temperature.

In the present work, we employ the model introduced in \cite{Polotsky:2012}
for a comprehensive study of the RC adsorption onto the RS beyond the
transition point in the annealed approximation. It is important to note that for the system considered,
 the annealed approximation is interesting not only as a mathematical trick but
due to its correspondence to real physical situations. Mathematically,
the annealed approximation is equivalent to direct averaging the partition function of
a disordered system instead of averaging its logarithm (i.e., free
energy) over all possible realizations of the disorder. For our system
this means that monomers and sites participate in thermal motion along
with the conformational degrees of freedom, and, following Grosberg
\cite{Grosberg:1985}, we may refer to the annealed RC (RS) as a copolymer
(a heterogeneous surface) with a ``mobile primary structure''.

Yoshinaga et al. \cite{Yoshinaga:2008} developed a theory of
adsorption of the so-called two-state polymers consisting of monomer
units that can change its type (i.e., its affinity to the substrate
or its hydrophobicity/hydro\-phi\-li\-ci\-ty). The authors of \cite{Yoshinaga:2008}
showed the equivalence of the two-state polymer to the annealed
two-letter RC and established the correspondence between the RC parameters
(\emph{a priori} probabilities to find a monomer of a certain type
in the RC sequence) and
standard chemical potentials of monomers
in the bulk and at the surface.
Another situation that can be described in terms of the annealed approximation was suggested
in \cite{Grosberg:1985}: the annealed RC can be viewed as a homopolymer
consisting of monomer units that can adsorb small molecules, like
surfactants \cite{Currie:2001}, from the solution. As a result,
there are two types of monomer units: free and with an adsorbed ``side
group''. Correspondingly, monomer units with ``side groups'' may have
additional attraction to each other or to a surface. This picture
can be straightforwardly extended to the case where two interacting
objects are heterogeneous: both polymer and surface may be ``two state''
or may bind (different) ligands and, being in a bound state, attract
each other.

There exists another aspect that makes the study of RC adsorption
onto RS in the annealed approximation interesting. As it is well known, in the case of
a RC adsorbing onto a homogeneous surface, the situation develops
as follows \cite{Orlandini:2002,Polotsky:2009}: with a decreasing temperature,
transformation of non-adsorbing monomer units into adsorbing ones
occurs. As a result, the \emph{a posteriori} first and higher moments
of the monomer distribution (that is, the number of monomer units of \emph{A}
and \emph{B} type, the number of dyads: \emph{AA}, \emph{BB}, \emph{AB}
and \emph{BA}, triads: \emph{AAA}, \emph{ABA}, \emph{BAA}, \dots  and
so on) do not correspond to their expectation values. In the case
where both interacting objects are heterogeneous and there are two
or more possibilities of forming favorable, or ``good'', contacts
(say, there are two kinds of favorable contacts: \emph{Aa} and \emph{Bb})
it is not easy to predict the system behavior. The most unclear will
be the symmetric situation, where both good contacts (\emph{Aa} and
\emph{Bb}) are equally favorable, whereas non-attractive ``bad'' contacts
(\emph{Ab} and \emph{Ba}) are equally unfavorable and, in addition,
both the RC and the RS have a symmetric composition (i.e., equal amounts
of \emph{A} and \emph{B} monomers and \emph{a} and \emph{b} sites):
here, one cannot say in advance how this compositional equilibrium
will be biased during the interaction of the RC with the RS upon a decreasing
temperature (increasing interaction strength). Phase diagram for the
symmetric case was analyzed in \cite{Polotsky:2012}: it was shown
that RC tends to adsorb onto the RS with the same type of correlations:
quasi-blocky on quasi-blocky, quasi-alternating on quasi-alternating
whereas uncorrelated, or Bernoullian, RCs (RSs) do not ``feel''
correlations on the RS (in the RC). These findings are in line with
the results of Monte Carlo simulations~\cite{Ziebarth:2008} for
a more realistic three-dimensional self-avoiding chain.

The rest of the paper is organized as follows. In section~\ref{sec:Model}, the model
and the formalism based on the annealed approximation and the GFs approach are introduced.
Presentation of the results is preceded by section~\ref{sec:refsys} where a simpler
system--a RC adsorbing onto a homogeneous surface--is considered to
illustrate the main features of the annealed approximation. This system also serves us
as a reference. Section~\ref{sec:result} is the main part of the paper devoted to
the exposition of the results in the case of symmetric composition and
symmetric ``interaction map'' (``interaction matrix''). We present
temperature dependences of the total and partial adsorbed fractions,
analyze the effect of a variation in the composition and in the character
of correlations in the RC and the RS, discuss the effect of asymmetry
in the composition and/or interaction matrix on the run of these dependences.
Finally, we summarize in section~\ref{sec:summary}.

\section{\label{sec:Model}Model and method}

\subsection{\label{sub:model_definition}Definition of the model}

\begin{figure}[ht]
\begin{center}
\includegraphics[width=7cm]{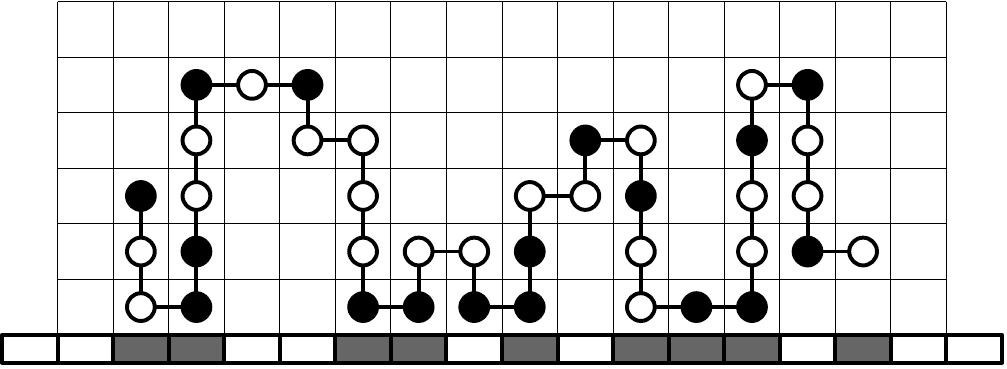}
\end{center}
\caption{\label{fig:Model} 2D partially directed walk model of random copolymer
consisting of {\em A} (black) and {\em B} (white) monomer units
near the linear (``planar'') random surface composed of {\em a} (gray)
and {\em b} (white) sites.}
\end{figure}

Consider a RC chain composed of {\em A} and {\em B} monomer
units interacting with a RS that carries {\em a} and {\em b}
sites, figure~\ref{fig:Model}
(note that the two-species heteropolymer chain is similar to the so-called hydrophobic-polar (HP) copolymer, a model widely used in theoretical studies of protein folding~\cite{Dill:1985, Lau:1989}).
As in \cite{Polotsky:2012}, we model
polymer conformations by  2D-PDWs on the square lattice, therefore,
the adsorbing surface is simply a line. The monomer sequence of the
RC is given by $\chi=\{\chi_{1},\chi_{2},\dots ,\chi_{n}\}$, where $\chi_{i}=A\mbox{ or }B$
while the surface pattern is given by $\sigma=\{\dots,\sigma_{1},\sigma_{2},\sigma_{3},\dots\}$,
where $\sigma_{x}=a\mbox{ or }b$.

Monomer units and surface sites are distributed randomly and modelled
as the  first order Markov chain. For the RC, the latter is determined
by the probabilities to find {\em A} and {\em B} units in the
sequence: $P(\chi_{i}=A)=f_{A}$ and $P(\chi_{i}=B)=f_{B}=1-f_{A}$,
and by the probabilities that the monomer of the type $i$ is followed
by the monomer of the type $j$, $P(\chi_{m}=j|\chi_{m-1}=i)=p_{ij}$.
The correlation parameter $c_{\mathrm{p}}=1-p_{AB}-p_{BA}=p_{AA}+p_{BB}-1$
determines the character of correlations in the sequence: $c_{\mathrm{p}}>0$
means that there is a tendency in the sequence for grouping similar
monomers into clusters, $c_{\mathrm{p}}<0$ favors the alternating sequence
of {\em A}'s and {\em B}'s, $c_{\mathrm{p}}=0$ corresponds to uncorrelated
(Bernoullian) sequences. The probability of occurrence of a particular
realization of the sequence $\chi$ is given by the product $P(\chi)=f_{\chi_{1}}p_{\chi_{1}\chi_{2}}p_{\chi_{2}\chi_{3}}\cdot\ldots\cdot p_{\chi_{n-1}\chi_{n}}$.

Similarly, the sequence of surface sites is determined by the probabilities
to find {\em a} and {\em b} site on the surface $P(\sigma_{x}=a)=g_{\mathrm{a}}\,\mbox{ and }\, P(\sigma_{x}=b)=g_{b}=1-g_{\mathrm{a}}$
and the probabilities that the site of the type $i$ is followed by
the site of the type $j$, $P(\sigma_{x}=j|\sigma_{x-1}=i)=s_{ij}$.
The correlation parameter for the surface $c_{\mathrm{s}}=1-s_{ab}-s_{ba}=s_{aa}+s_{bb}-1$.
The probability of a particular realization of the sequence $\sigma=\{\sigma_{1},\sigma_{2},\dots ,\sigma_{\ell}\}$
is given by the product $P(\sigma)=g_{\sigma_{1}}s_{\sigma_{1}\sigma_{2}}s_{\sigma_{2}\sigma_{3}}\cdot\ldots\cdot s_{\sigma_{\ell-1}\sigma_{\ell}}$.

The Hamiltonian (the energy) of the system can be written as follows:
\begin{equation}
H=\sum_{i=1}^{N}\delta_{y_{i},0}\,\epsilon_{\chi_{i}\sigma_{x_{i}}}\,,\label{eq:Hamiltonian}
\end{equation}
and the partition function of the system for particular realizations of
the RC and the RS is given by
\begin{equation}
Z_{n}(\beta|\chi,\sigma)=\sum_{\mathbf{r}_{i}}\exp\left[-\beta\sum_{i=1}^{N}\delta_{y_{i},0}\,\epsilon_{\chi_{i}\sigma_{x_{i}}}\right],\label{eq:Z_general_1}
\end{equation}
where $\mathbf{r}_{i}=\{x_{i},\: y_{i}\}$, $i=1,\dots,n$ denotes
the chain conformation (the trajectory of the chain), $\delta_{ij}$ is the Kronecker delta,
$\epsilon_{ij}$ is the energy of the monomer-surface contact ($i\in\{A,\, B\}$
and $j\in\{a,\, b\}$), and $\beta=1/k_{\mathrm{B}}T$ is the inverse temperature.

\subsection{Annealed approximation and generating functions approach }

We will solve the problem in the annealed approximation, where the quenched free energy
$\beta F_{\mathrm{q}}=-\langle\langle\ln Z_{n}(\beta|\chi,\sigma)\rangle_{\chi}\rangle_{\sigma}$
obtained by averaging the logarithm of the partition function over
all possible realizations of the RC and the RS is approximated by
the annealed free energy $\beta F_{\mathrm{a}}=-\ln\langle\langle Z_{n}(\beta|\chi,\sigma)\rangle_{\chi}\rangle_{\sigma}$,
where the partition function is averaged prior to taking the logarithm.
Here, the angular brackets $\left\langle \ldots\right\rangle $ denote
averaging over sequence or surface randomness.

To calculate the annealed partition function of the  system considered,
$\langle\langle Z_{n}(\beta|\chi,\sigma)\rangle_{\chi}\rangle_{\sigma}$,
we use the generating functions (GFs) approach (or the grand canonical
approach). In the case of adsorption onto heterogeneous surface, this
approach consists in calculating the GF
\begin{equation}
\Xi(z,t)=\sum_{n=1}^{\infty}\sum_{m=1}^{n}\langle\langle Z_{n,m}(\beta|\chi,\sigma)\rangle_{\chi}\rangle_{\sigma}z^{n}t^{m}\,,\label{eq:GF_def}
\end{equation}
 where $Z_{n,m}(\beta|\chi,\sigma)$ is the constrained partition
function of a chain with $n$ monomer units and the length of the
chain projection onto adsorbing substrate equal to $m$. The GF variables
$z$ and $t$ conjugate to the chain length and the chain projection,
respectively. The smallest singularity $z_c(t)$ of $\Xi(z,t)$ calculated at
$t=1$ gives an asymptotic expression for the partition function
in the large $n$ limit: $\langle\langle Z_{n}(\beta|\chi,\sigma)\rangle_{\chi}\rangle_{\sigma}\simeq z_{\mathrm{c}}^{-n}(t=1)$.
Then, the monomer chemical potential (the free energy per monomer unit)
$\mu=\ln[z_{\mathrm{c}}(t=1)]$

As it was shown in \cite{Polotsky:2012}, the smallest singularity
of $\Xi(z,t)$ is associated with the smallest root of the equation
\begin{equation}
\det[\mathbf{E}-\mathbf{\Xi}_{\mathrm{L}}(z,t)\mathbf{\Xi}_{\mathrm{S}}(zt)]=0.\label{eq:det_ann}
\end{equation}
In equation (\ref{eq:det_ann}), the functions $\mathbf{\Xi}_{\mathrm{S}}(zt)$,
$\mathbf{\Xi}_{\mathrm{L}}(z,t)$ are the GFs of adsorbed segments (usually
called ``trains'') and loops, respectively, in the matrix form:
\begin{eqnarray}
 &  & \mathbf{\Xi}_{\mathrm{S}}(zt)=\sum_{n=1}^{\infty}\Omega_{\mathrm{S}}(n)(zt)^{n}\mathbf{R}{}^{n}\,,\nonumber \\
 &  & \mathbf{\Xi}_{\mathrm{L}}(z,t)=\sum_{n=2}^{\infty}\sum_{m=2}^{n}\Omega_{\mathrm{L}}(n,m)z^{n}t^{m-2}\,(\mathbf{P}^{n}\otimes\mathbf{S}^{m-2}),\label{eq:GF_SLT}
\end{eqnarray}
 where $\Omega_{\mathrm{L}}(n,m)$ is the number of loops of contour length
$n$ and projection length $m$ and $\Omega_{\mathrm{S}}(n,m)$ is the number
of trains of length $n$. The matrices $\mathbf{P}$ and $\mathbf{S}$
are the transition probability matrices for RC and RS, respectively:
\begin{eqnarray}
\mathbf{P} & = & \left(\begin{array}{cc}
p_{AA} & p_{AB}\\
p_{BA} & p_{BB}
\end{array}\right)=\left(\begin{array}{cc}
p_{AA} & 1-p_{AA}\\
1-p_{BB} & p_{BB}
\end{array}\right),\nonumber \\
\mathbf{S} & = & \left(\begin{array}{cc}
s_{aa} & s_{ab}\\
s_{ba} & s_{bb}
\end{array}\right)=\left(\begin{array}{cc}
s_{aa} & 1-s_{aa}\\
1-s_{bb} & s_{bb}
\end{array}\right),
\end{eqnarray}
$\mathbf{P}\otimes\mathbf{S}$ is the Kronecker product of the these
matrices. The matrix $\mathbf{R}$ is defined as
\begin{equation}
\mathbf{R}=(\mathbf{P}\otimes\mathbf{S})\cdot\mathbf{W}\label{eq:R_def}
\end{equation}
 with the diagonal ``interaction matrix''
\begin{equation}
\mathbf{W}=\diag(w_{Aa},\, w_{Ab},\, w_{Ba},\, w_{Bb})\label{eq:W_def}
\end{equation}
containing all statistical weights of different monomer-surface contacts, $w_{ij}\equiv \re^{-\beta\epsilon_{ij}}$,
where $i\in\{A,\, B\}$ and $j\in\{a,\, b\}$.

The matrix GFs $\mathbf{\Xi}_{\mathrm{S}}(zt)$ and $\mathbf{\Xi}_{\mathrm{L}}(z,t)$
can be easily calculated by using the eigenvalues and eigenvectors
of the matrices $\mathbf{P}$, $\mathbf{S}$, and $\mathbf{R}$ and
the expressions for scalar GFs of trains calculated straightforwardly
as follows:
\begin{equation}
\Xi_{\mathrm{S}}(z)=\sum_{n=1}^{\infty}\Omega_{\mathrm{S}}(n)z^{n}=z^{2}+z^{3}+\dots=\frac{z^{2}}{1-z}\,,
\end{equation}
and loops (calculated by using a loop decomposition
described in \cite{Polotsky:2012})
\begin{equation}
\Xi_{\mathrm{L}}(z,t)=\sum_{n=2}^{\infty}\sum_{m=2}^{n}\Omega_{\mathrm{L}} (n,m)z^{n}t^{m-2}=\frac{1-zt-z^{2}-z^{3}t-\sqrt{(1-zt-z^{2}-z^{3}t)^{2}-4z^{4}t^{2}}}{2z^{2}t^{2}}\,,
\end{equation}
 see also~\cite{Polotsky:2012}, equations~(23)--(25) for details.

Note that the determinant equation (\ref{eq:det_ann}) is a generalization
of the analogous scalar equation for a homopolymer adsorption onto
a homogeneous surface~ \cite{Birshtein:1979}: $1-\Xi_{\mathrm{L}}(z)\Xi_{\mathrm{S}}(wz)=0$,
where $w=\re^{-\beta\epsilon}$ is the statistical weight of a monomer-surface
contact, $\Xi_{\mathrm{L}}(z)=\Xi_{\mathrm{L}}(z,1)$.

The smallest singularity $z_{\mathrm{c}}$ must then be compared with the smallest
singularity $z_{\mathrm{V}}$ of the GF for the free (desorbed) chain in a
bulk $\Xi_{\mathrm{V}}(z)=\sum_{n=1}^{\infty}\Omega_{\mathrm{V}}(n)z^{n}$, which does
not depend on monomer sequence. Here, $\Omega_{\mathrm{V}}(n)$ is the number
of conformation that a chain of $n$ monomer units can acquire in the bulk. For the 2D-PDW polymer model $z_{\mathrm{V}}=\sqrt{2}-1$
\cite{Privman:1989}.

In the adsorption transition point $z_{\mathrm{c}}=z_{\mathrm{V}}$, hence, the equation
\begin{equation}
\det\left[\mathbf{E}-\mathbf{\Xi}_{\mathrm{L}}(z_{\mathrm{V}},t)\mathbf{\Xi}_{\mathrm{S}}(z_{\mathrm{V}}t)\right]_{t=1}=0\label{eq:tp}
\end{equation}
 determines the position of the transition point.

\subsection{\label{sub:observables}Calculation of observables}

Various observables can be found via differentiation of the smallest
singularity $z_{\mathrm{c}}(t)$ of the GF. For example, logarithmic derivative
of $z_{\mathrm{c}}$ with respect to the statistical weight of different monomer-surface
contacts, $w_{ij}$, gives the average fraction of these contacts
occurring in the adsorbed RC chain:
\begin{equation}
\theta_{ij}=-\frac{\partial\ln z_{\mathrm{c}}}{\partial\ln w_{ij}}=-\frac{w_{ij}}{z_{\mathrm{c}}}\cdot\frac{\partial z_{\mathrm{c}}}{\partial w_{ij}}\label{eq:qij}\,.
\end{equation}
 The total adsorbed fraction is given by the sum of four contributions:
$\theta=\theta_{Aa}+\theta_{Ab}+\theta_{Ba}+\theta_{Bb}$.

The partial derivative $\partial z_{\mathrm{c}}/\partial w_{ij}$ in equation
(\ref{eq:qij}) can be calculated by differentiating the equation
(\ref{eq:det_ann}). Namely, if we denote in the left-hand side of
equation (\ref{eq:det_ann}) as $D:=\det[\mathbf{E}-\mathbf{\Xi}_{\mathrm{L}}(z,\, t)\mathbf{\Xi}_{\mathrm{S}}(zt)]$,
then
\begin{equation}
\frac{\partial z_{\mathrm{c}}}{\partial w_{ij}}=-\left.\frac{\partial D/\partial w_{ij}}{\partial D/\partial z}\right|_{z=z_{\mathrm{c}},\: t=1}\,.
\end{equation}
 The derivatives of $D$ can be found with the aid of Jacobi's formula
\begin{equation}
\frac{\rd\det\mathbf{X}(\tau)}{\rd\tau}=\Tr\left(\Adj(\mathbf{X})\cdot\frac{\rd\mathbf{X}}{\rd\tau}\right)\,,
\end{equation}
 where ``Tr'' stands for the trace of a matrix and $\Adj(\mathbf{X})$
denotes the adjugate matrix for $\mathbf{X}$.

By analogy, one can calculate other observables by choosing a proper
differentiation variable. Taking a derivative of $z_{\mathrm{c}}(t)$ with
respect to $t$ gives access to the ratio of the average projection
of the RC chain onto the surface to the RC contour length:
\begin{equation}
\frac{\left\langle m\right\rangle }{n}=-\left.\frac{\partial\ln z_{\mathrm{c}}}{\partial\ln t}\right|_{t=1}=-\frac{1}{z_{\mathrm{c}}}\cdot\left.\frac{\partial z_{\mathrm{c}}}{\partial t}\right|_{t=1}\,.
\end{equation}

Since in the annealed system, RC monomer units and RS sites are in
thermal motion, it is especially interesting to find equilibrium moments
of their distributions. To calculate the fraction of \emph{A} monomers
in the RC chain, $\nu_{A}$, let us introduce an auxiliary term into
the Hamiltonian (\ref{eq:Hamiltonian}): we set $H\to H+\Delta H$
, where
\begin{equation}
\Delta H=-(\eta/\beta)\sum_{i=1}^{N}\chi_{i}\,.
\end{equation}
Later $\eta$ will be set to zero. With this auxiliary term, the matrix
$\mathbf{P}$ modifies as follows:
\begin{equation}
\mathbf{P}=\left(\begin{array}{cc}
p_{AA}\re^{\eta} & p_{AB}\\
p_{BA}\re^{\eta} & p_{BB}
\end{array}\right).
\end{equation}
The matrix $\mathbf{R}$ is again $\mathbf{R}=(\mathbf{P}\otimes\mathbf{S})\cdot\mathbf{W}$.
Then, the average fraction of \emph{A} monomers in the RC chain is
given by
\begin{equation}
\nu_{A}=-\left.\frac{\partial\ln z_{\mathrm{c}}}{\partial\eta}\right|_{\eta=0}=-\left.\frac{1}{z_{\mathrm{c}}}\cdot\frac{\partial z_{\mathrm{c}}}{\partial\eta}\right|_{\eta=0}\label{eq:nu_a_p}\,.
\end{equation}
The fraction of \emph{B} units follows automatically: $\nu_{B}=1-\nu_{A}$.

Similarly, one can calculate the average fraction of \emph{AA}, \emph{AB},
\emph{BA}, or \emph{BB} dyads in the sequence. For example, $\nu_{AA}$
is calculated by introducing the following auxiliary term:
\begin{equation}
\Delta H=-(\eta/\beta)\sum_{i=2}^{N}\chi_{i-1}\chi_{i}\,.
\end{equation}
This modifies the matrix $\mathbf{P}$ in the following way:
\begin{equation}
\mathbf{P}=\left(\begin{array}{cc}
p_{AA}\re^{\eta} & p_{AB}\\
p_{BA} & p_{BB}
\end{array}\right),
\end{equation}
i.e., the multiplier $\re^{\eta}$ appears at the corresponding element
$(p_{AA})$ of the probability matrix. The cluster parameter $\lambda_{\mathrm{p}}$,
which is the \emph{a posteriori} analogue of the \emph{a priori} correlation
parameter $c_{\mathrm{p}}$, is calculated as $\lambda_{\mathrm{p}}=1-\nu_{AB}/\nu_{A}-\nu_{BA}/\nu_{B}=\nu_{AA}/\nu_{A}-\nu_{BB}/\nu_{B}-1$.

In a similar manner, we can use the same idea to obtain the
composition of RS. Thus, to calculate the fraction of \emph{a} sites
on the surface, the matrix $\mathbf{S}$ should be modified as follows:
\begin{equation}
\mathbf{S}=\left(\begin{array}{cc}
s_{aa}\re^{\eta} & s_{ab}\\
s_{ba}\re^{\eta} & s_{bb}
\end{array}\right).
\end{equation}

However, there is an important difference between RC and RS: while
in the former case all monomer units can be involved in the interaction
with the surface, in the latter case only a part of the RS may be
involved in the interaction with the polymer chain, the size of this
``contact zone'' is equal to the projection of the RC on the substrate
$m\leqslant  n$. Mathematically, this difference is expressed in the limits
of summation in the GF of equation (\ref{eq:GF_def}). In this sense,
the quantity ``average fraction of \emph{a} sites on the surface''
is not well defined for the whole surface because if one takes a finite
but large surface which cannot be completely covered by the RC there
will be two parts of the surface, i.e., a part involved and a part not involved
into interaction with the RC. A better quantity is the ``local'' fraction
of \emph{a}-sites, i.e., the ratio of the number of \emph{a} sites in the
``contact zone'', $m_{\mathrm{a}}$ to the (instantaneous) size of this zone
$m$.

An expression analogous to equation (\ref{eq:nu_a_p}) does not give the
sought fraction of \emph{a} sites on the surface. It gives the ratio of
the average number of \emph{a} sites on the surface ``occupied'' by
the RC, $m_{\mathrm{a}}$, to the contour length i.e. $m_{\mathrm{a}}/n$. Therefore,
a correct estimate of the ``local'' fraction of \emph{a}-sites,
$\nu_{\mathrm{a}}$, will be given by
\begin{equation}
\nu_{\mathrm{a}}=-\left.\frac{\partial\ln z_{\mathrm{c}}}{\partial\eta}\right|_{\eta=0}\left(\frac{\left\langle m\right\rangle }{n}\right)^{-1}=-\left.\frac{1}{z_{\mathrm{c}}}\cdot\frac{\partial z_{\mathrm{c}}}{\partial\eta}\right|_{\eta=0}\left(\frac{\left\langle m\right\rangle }{n}\right)^{-1}
\end{equation}
and similarly for the number of dyads. The cluster parameter for the
RS is calculated as $\lambda_{\mathrm{s}}=1-\nu_{ab}/\nu_{\mathrm{a}}-\nu_{ba}/\nu_{b}$.

\section{\label{sec:refsys}Reference system: RC adsorbed on homogeneous substrate}

In the Introduction we briefly discussed the merits and demerits and
the fields of application of the annealed approximation in the study of (various) disordered
systems. It will be instructive to illustrate some peculiarities of
the annealed approximation in the case of a simpler system, where a RC is adsorbed onto
a homogeneous surface. This problem was studied earlier: while in
\cite{Orlandini:2002}, the Bernoullian RC was considered, in \cite{Polotsky:2009}
a general program for correlated Markovian RC in the framework of
the GF approach on the lattice was developed. The results in \cite{Polotsky:2009}
were obtained for a simple random walk model of polymer in three
dimensions, although the approach is quite general and may be straightforwardly
extended to other types of lattice models of polymers or to other
geometries of adsorbing substrates. In the present work, we apply this
program for the 2D-PDW model used in the present study. This will
also serve as a ``reference system'' for comparison
with our ``original'' system.

As it was shown \cite{Polotsky:2009}, for a RC adsorbed onto a strictly
homogeneous surface, the smallest singularity of the GF $\Xi(z)=\sum_{n=1}^{\infty}\langle Z_{n}(\beta|\chi)\rangle_{\chi}z^{n}$
in the annealed approximation is found from the equation similar to equation (\ref{eq:det_ann}):
$\det[\mathbf{E}-\mathbf{\Xi}_{\mathrm{L}}(z)\mathbf{\Xi}_{\mathrm{S}}(z)]=0$, with
matrix GFs $\mathbf{\Xi}_{\mathrm{L}}(z)=\sum_{n=2}^{\infty}\Omega_{\mathrm{L}}(n)z^{n}\,\mathbf{P}^{n}$
and $\mathbf{\Xi}_{\mathrm{S}}(z)=\sum_{n=2}^{\infty}\Omega_{\mathrm{S}}(n)z^{n}\,\mathbf{R}^{n}$,
$\mathbf{R}=\mathbf{P}\cdot\mathbf{W}$, where $\mathbf{W}=\diag(w_{A},\, w_{B})=\diag(\re^{-\beta\epsilon_{A}},\, \re^{-\beta\epsilon_{B}})$
is the diagonal matrix of statistical weights of \emph{A} and \emph{B}
contacts with the surface. Calculation of observables is similar to
that described is section~\ref{sub:observables}. For more details see
\cite{Polotsky:2009}.

These formulas can also be directly obtained from equations~(\ref{eq:GF_def})--(\ref{eq:W_def})
by assuming that the RS contains only \emph{a} sites and, correspondingly,
by replacing the matrix $\mathbf{S}$ in equations (\ref{eq:GF_def})--(\ref{eq:R_def})
by $1\times1$ unity matrix, $\mathbf{S}=1$, redefining $\epsilon_{Aa}\equiv\epsilon_{A}$
and $\epsilon_{Ba}\equiv\epsilon_{B}$, and setting $t=1$. Then, the
interaction matrices $\mathbf{W}$, equation (\ref{eq:W_def}), and
$\mathbf{R}$, equation (\ref{eq:R_def}) will have the dimensionality
$2\times2$. Alternatively, one can consider \emph{a} and \emph{b}
sites as identical, i.e., set $\epsilon_{Aa}=\epsilon_{Ab}=\epsilon_{A}$
and $\epsilon_{Ba}=\epsilon_{Bb}=\epsilon_{B}$ and then directly
use the formalism introduced in section~\ref{sec:Model}.

We consider the case of adsorbing \emph{A} and neutral \emph{B} contacts:
$\epsilon_{A}=-1$, $\epsilon_{B}=0$. Without loss of generality,
let us assume that \emph{A} and \emph{B} monomers have equal probabilities
to be found in the monomer sequence, $f_{A}=f_{B}=0.5$ but have various
character of correlations in the RC chain (i.e., various $c_{\mathrm{p}}$).
For the sake of comparison, we also consider the homopolymer (HP) consisting
of adsorbing \emph{A} monomer units only.

\begin{figure}[ht]
\includegraphics[width=4.5cm]{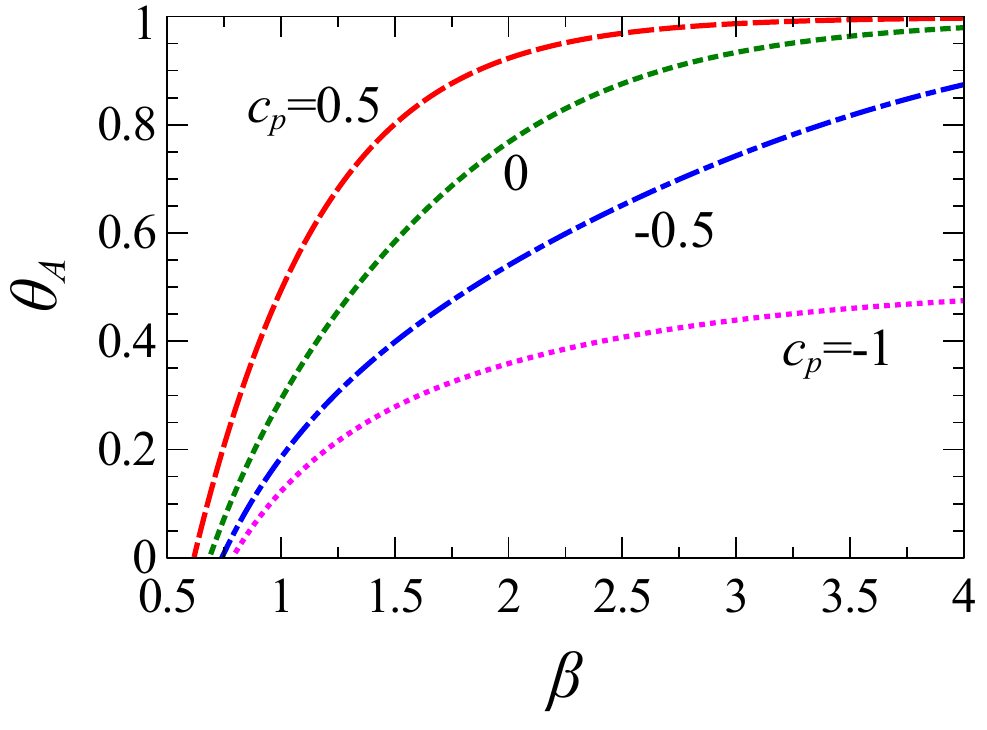}
\includegraphics[width=4.5cm]{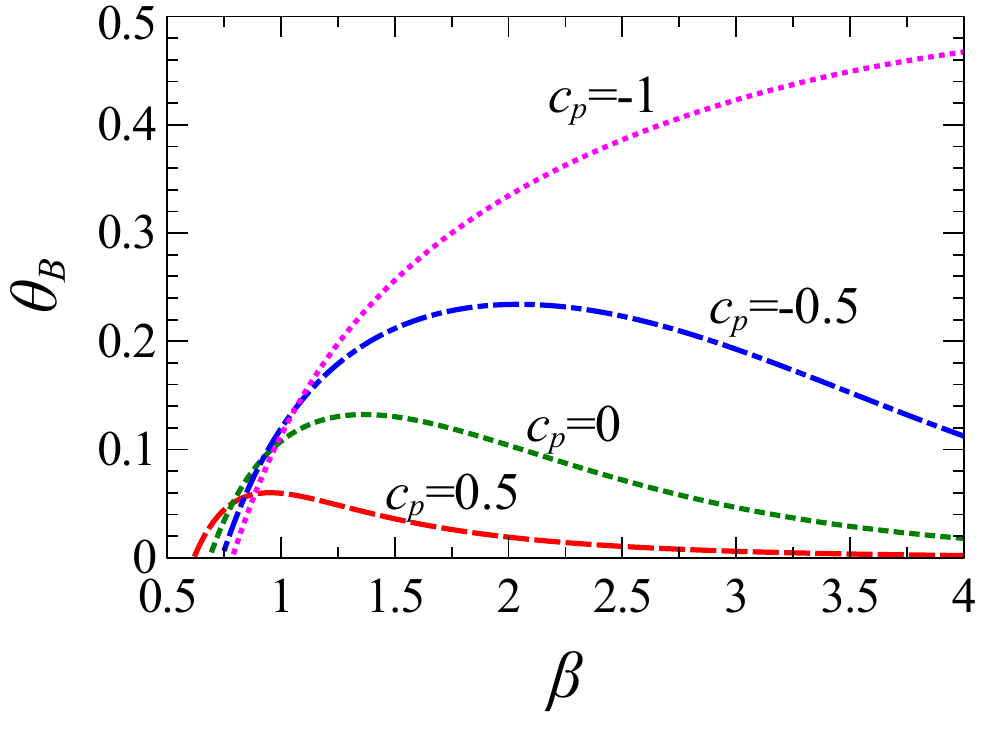}
\includegraphics[width=4.5cm]{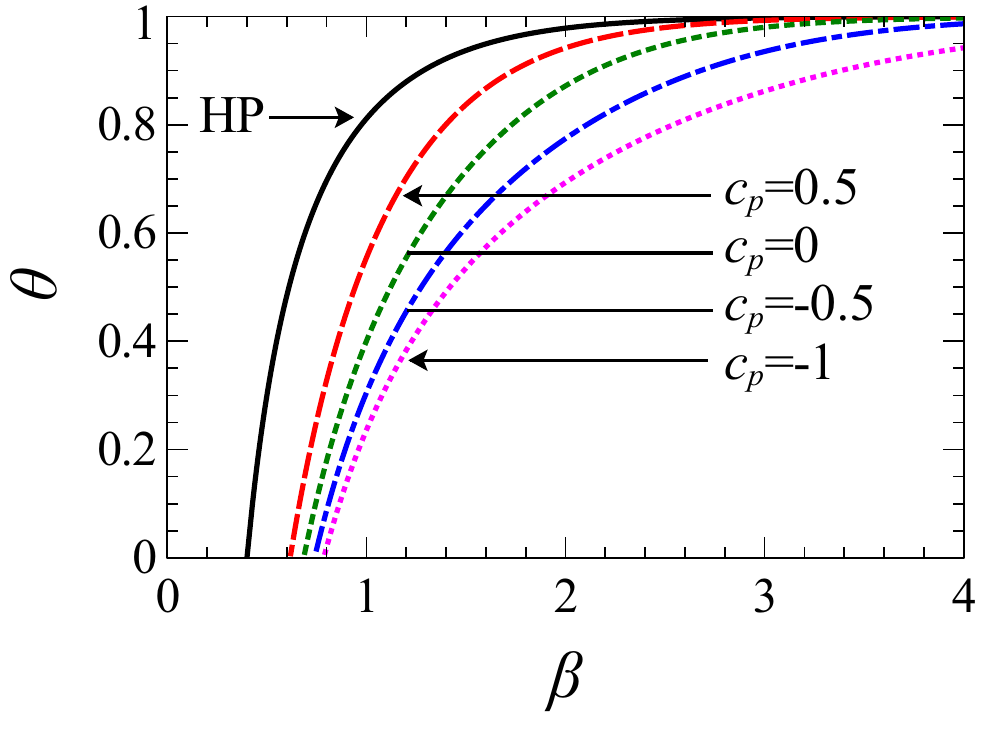}
\\%
\parbox[t]{0.32\textwidth}{%
\centerline{(a)}%
}%
\hfill%
\parbox[t]{0.32\textwidth}{%
\centerline{(b)}%
}%
\hfill
\parbox[t]{0.32\textwidth}{%
\centerline{(c)}%
}%
\hspace{5mm}
\\
\begin{center}
\includegraphics[width=4.5cm]{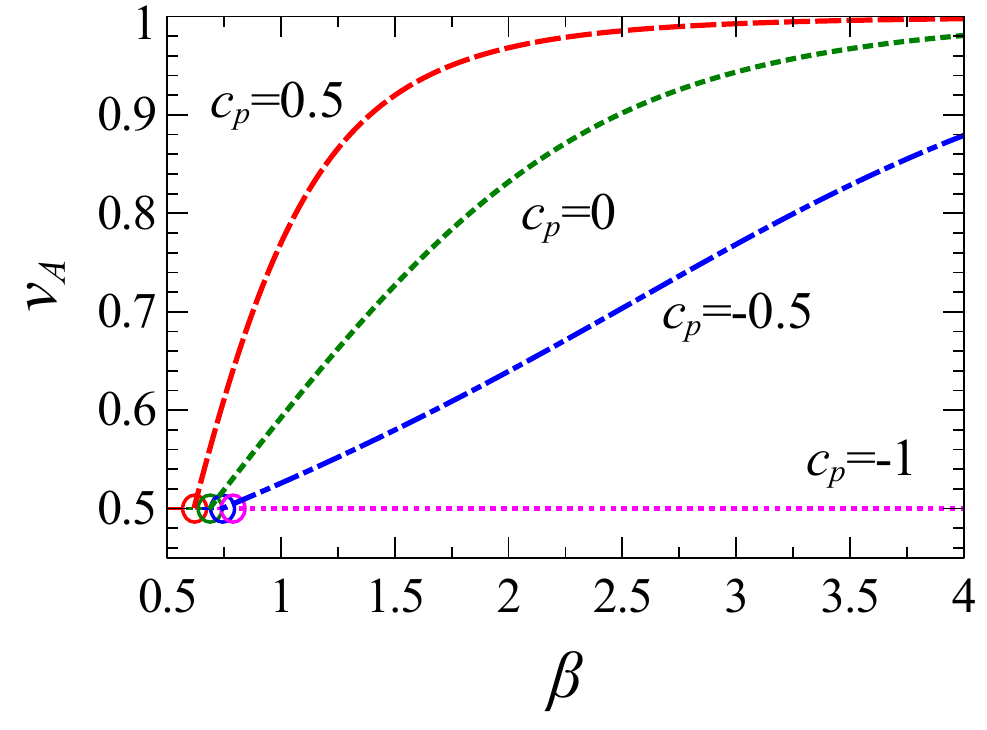}
\hspace{1cm}
\includegraphics[width=4.5cm]{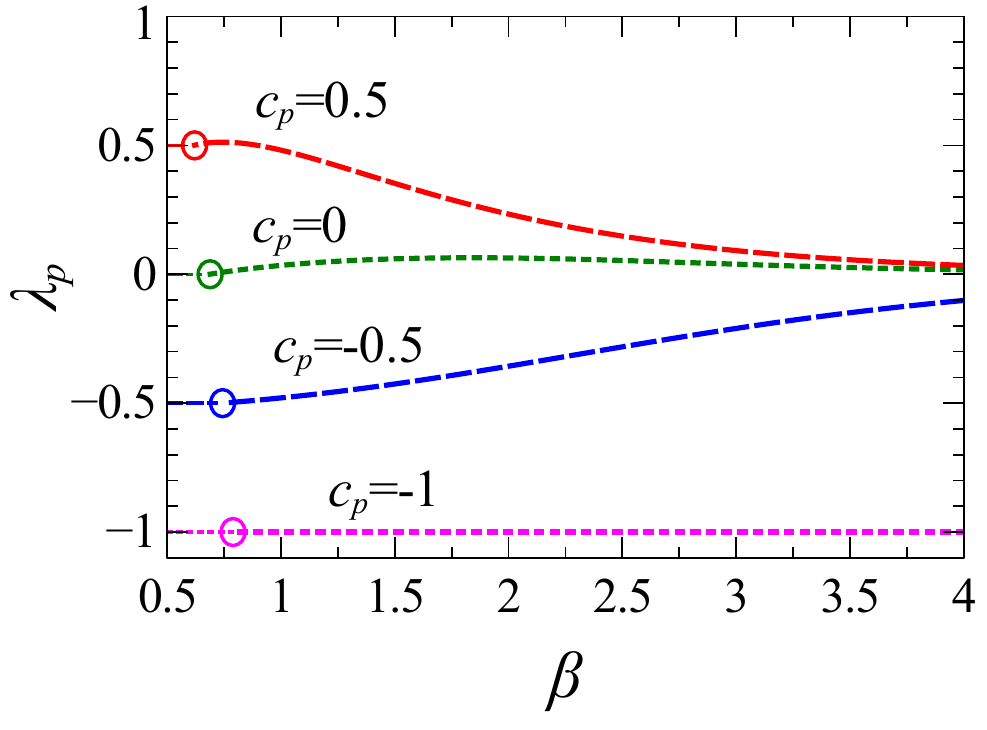}
\end{center}
\begin{center}
\parbox[t]{0.32\textwidth}{%
\centerline{(d)}%
}%
\hspace{1cm}%
\parbox[t]{0.32\textwidth}{%
\centerline{(e)}%
}%
\end{center}
\caption{\label{fig:refsys}(Color online) Fraction of \emph{A}- (good) (a) and \emph{B}-
(bad) (b) contacts, total adsorbed fraction (c), average fraction
of \emph{A}-monomers, and (d) cluster parameter for RC (e) in the
reference system as functions of inverse temperature $\beta$ calculated
for $f_{A}=0.5$ and various $c_{\mathrm{p}}$ as indicated.}
\end{figure}
Figure~\ref{fig:refsys} shows temperature dependences of contact
fractions (a, b), total adsorbed amount (c), fraction of \emph{A} monomers
$\nu_{A}$ (d), and cluster parameter $\lambda_{\mathrm{p}}$ (e).  At $c_{\mathrm{p}}>-1$, with an increasing $\beta$, the number of adsorbed\emph{
A} monomers as well as the overall adsorbed fraction monotonously
increase while the fraction of adsorbed \emph{B} units behaves non-monotonously
and vanishes at very high $\beta$. This is accompanied by interconversion
of repelling monomer units \emph{B} into adsorbing units \emph{A}
(however, at $\beta\leqslant \beta_{\mathrm{tr}}$, where $\beta_{\mathrm{tr}}$ is the transition
point, $\nu_{A}=f_{A}$ and $\lambda_{\mathrm{p}}=c_{\mathrm{p}}$) which explains the
observed depencences of $\theta_{A}$ and $\theta_{B}$. Only in the
case $c_{\mathrm{p}}=-1$, corresponding to the regularly alternating \emph{AB}-copolymer
its regular (i.e., quenched) sequence does not change and both $\theta_{A}$
and $\theta_{B}$ grow with increasing $\beta$ due to a cooperative
effect. Therefore, the annealed approximation corresponds
to the physical situation that essentially differs from the quenched
case.

\section{\label{sec:result}Results and discussion}

\subsection{Choosing the system parameters}

The system we consider depends on a large number of variables (to
be precise, there are 9 parameters: $f_{A}$, $g_{\mathrm{a}}$, $c_{\mathrm{p}}$,
$c_{\mathrm{s}}$, $\epsilon_{Aa}$, $\epsilon_{Ab}$, $\epsilon_{Ba}$, $\epsilon_{Bb}$,
and $\beta$). In \cite{Polotsky:2012} we have suggested a reasonable
choice of parameters by keeping the inverse temperature $\beta=1/k_{\mathrm{B}}T$
as a separate control variable and by fixing attractive and repulsive
energies, thus obtaining physically relevant temperature dependences.
In \cite{Polotsky:2012}, three different ``interaction schemes'' for
the monomer-site interaction energies were considered, the choice
was also motivated by the works of other authors.

Since the present work is devoted to the study of a particular case
of the RC and RS symmetric with respect both to composition, $f_{A}=g_{\mathrm{a}}=0.5$,
and monomer-site interactions, $\epsilon_{Aa}=\epsilon_{Bb}$ and
$\epsilon_{Ab}=\epsilon_{Ba}$, this restricts even more the set
of variable parameters; in fact, there remain only three ones: $c_{\mathrm{p}}$,
$c_{\mathrm{s}}$, and $\beta$. Therefore, the main part of the present work
will be devoted to the study of the symmetric (but highly non-trivial)
case;  the effect of the asymmetry on
the RC and the RS adsorption behaviour will be briefly discussed in the end of this section.
We will study the system with the following set of parameters: {\em
Aa} and {\em Bb} contacts are favorable, $\epsilon_{Aa}=\epsilon_{Bb}=-1$,
{\em Ab} and {\em Ba} contacts are neutral, $\epsilon_{Ab}=\epsilon_{Ba}=0$.

\subsection{Phase diagram and symmetry properties}

\begin{wrapfigure}{i}{0.5\textwidth}
\centerline{\includegraphics[width=8.1cm]{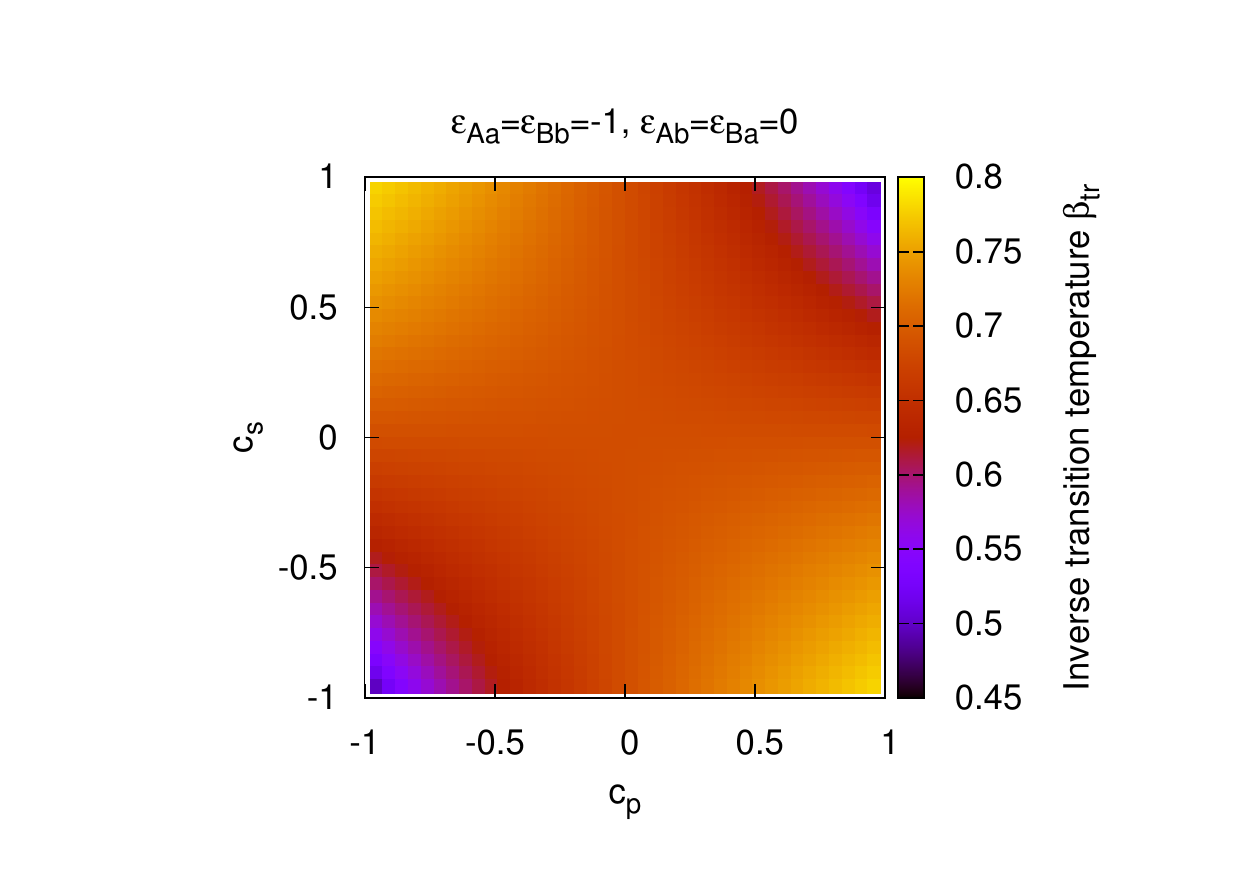}}
\caption{\label{fig:phd}(Color online) Inverse transition temperature in the annealed approximation
as function of RC and RS correlation parameters for $\epsilon_{Ab}=\epsilon_{Ba}=-1$,
$\epsilon_{Ab}=\epsilon_{Ba}=0$ in the symmetric case $f_{A}=g_{\mathrm{a}}=0.5$. }
\end{wrapfigure}
In \cite{Polotsky:2012}, phase diagram, i.e., the dependence of the inverse
adsorption transition temperature $\beta_{\mathrm{tr}}$ on the correlation
parameters $c_{\mathrm{p}}$ and $c_{\mathrm{s}}$, was calculated. Figure~\ref{fig:phd}
shows this diagram as a density plot in $(c_{\mathrm{p}},\, c_{\mathrm{s}})$ coordinates
(in such a form it was not presented in \cite{Polotsky:2012}).
As it follows from the diagram, the smallest values of $\beta_{\mathrm{tr}}$
(dark color, the bottom of the color scale in figure~\ref{fig:phd})
are observed in the vicinity of $(c_{\mathrm{p}},\, c_{\mathrm{s}})\approx(1,\,1)$
and $(-1,\,-1)$ whereas the largest values of $\beta_{\mathrm{tr}}$ (light
color, the top of the color scale in figure~\ref{fig:phd}) are observed in the
vicinity of $(c_{\mathrm{p}},\, c_{\mathrm{s}})\approx(-1,\,1)$ and $(1,\,-1)$. We
can also observe that this density plot is symmetric with respect
to the origin $(c_{\mathrm{p}},\, c_{\mathrm{s}})=(0,\,0).$ This means that $\beta_{\mathrm{tr}}(c_{\mathrm{p}}=x,\, c_{\mathrm{s}}=y)=\beta_{\mathrm{tr}}(c_{\mathrm{p}}=-x,\, c_{\mathrm{s}}=-y)$
where $x$ and $y$ may take on any value in the interval $(-1,\,1)$.
This symmetry of the diagram is the consequence of the interaction
and composition symmetries. (At a glance it may also seem that the diagram
is also symmetric with respect to the diagonals $c_{\mathrm{s}}=\pm c_{\mathrm{p}}$
but this is not the case).

This symmetry manifests itself in temperature dependences of the observables
beyond the transition point. We vary the values for the correlation
parameter for the RC and the RS considering the cases $c_{\mathrm{p}}=-0.5$,
0, 0.5, and $c_{\mathrm{s}}=-1$, $-0.5$, 0, 0.5. It is clear that various combinations
of $c_{\mathrm{p}}$ and $c_{\mathrm{s}}$ are possible ($3\times4=12$ combinations).
However, due to the interaction and the composition symmetries,
it turns out that these 12 dependences for an adsorbed fraction can be
presented on the same plot by 5 curves.

\subsection{Adsorbed fraction} % and thermodynamic quantities

\begin{figure}[ht]
\centerline{
\includegraphics[width=0.32\textwidth]{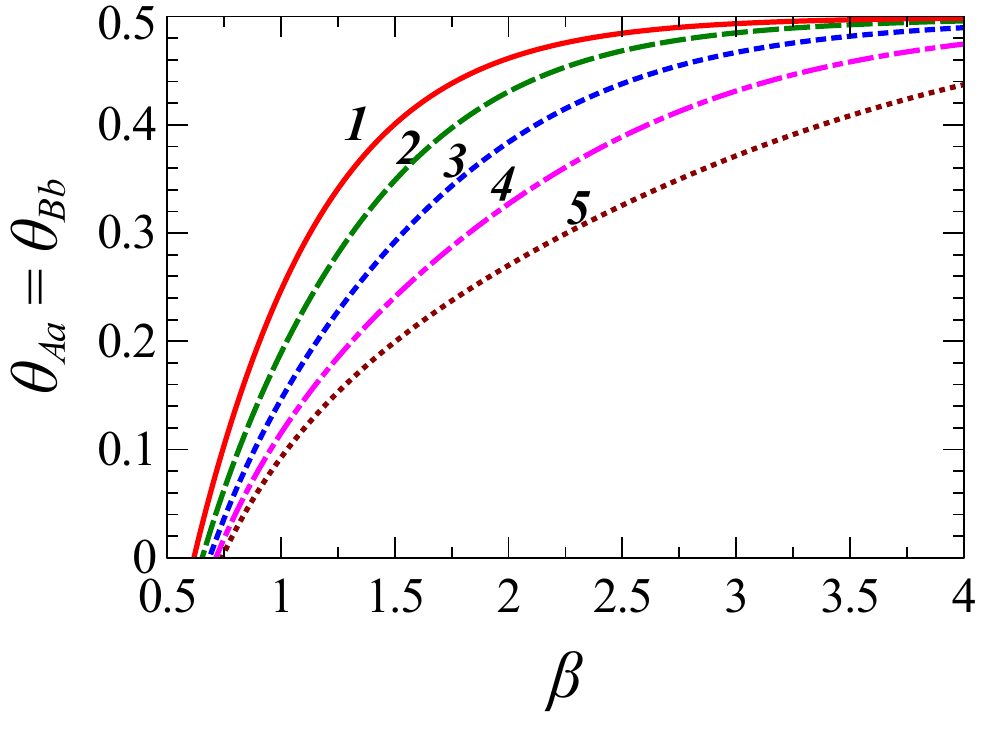}%
\hspace{2cm}%
\includegraphics[width=0.32\textwidth]{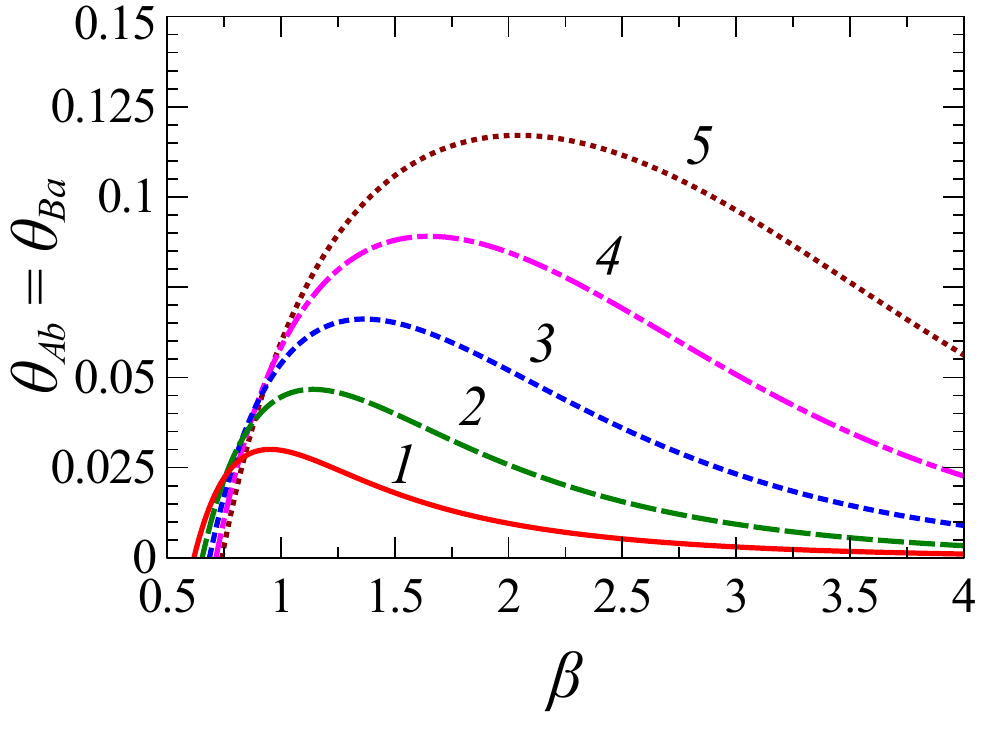}}%
%\\%
\begin{center}
\parbox[t]{0.32\textwidth}{%
\centerline{(a)}%
}%
\hspace{2cm}%
\parbox[t]{0.32\textwidth}{%
\centerline{(b)}%
}%
\end{center}
\centerline{
\includegraphics[width=0.32\textwidth]{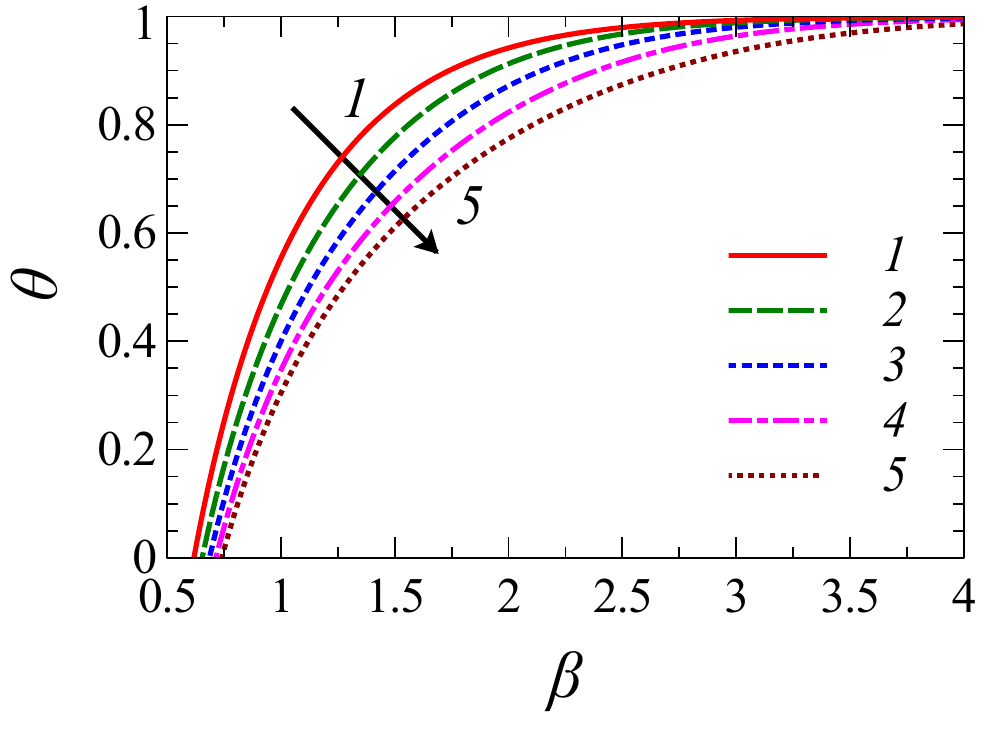}%
\hspace{2cm}%
\includegraphics[width=0.32\textwidth]{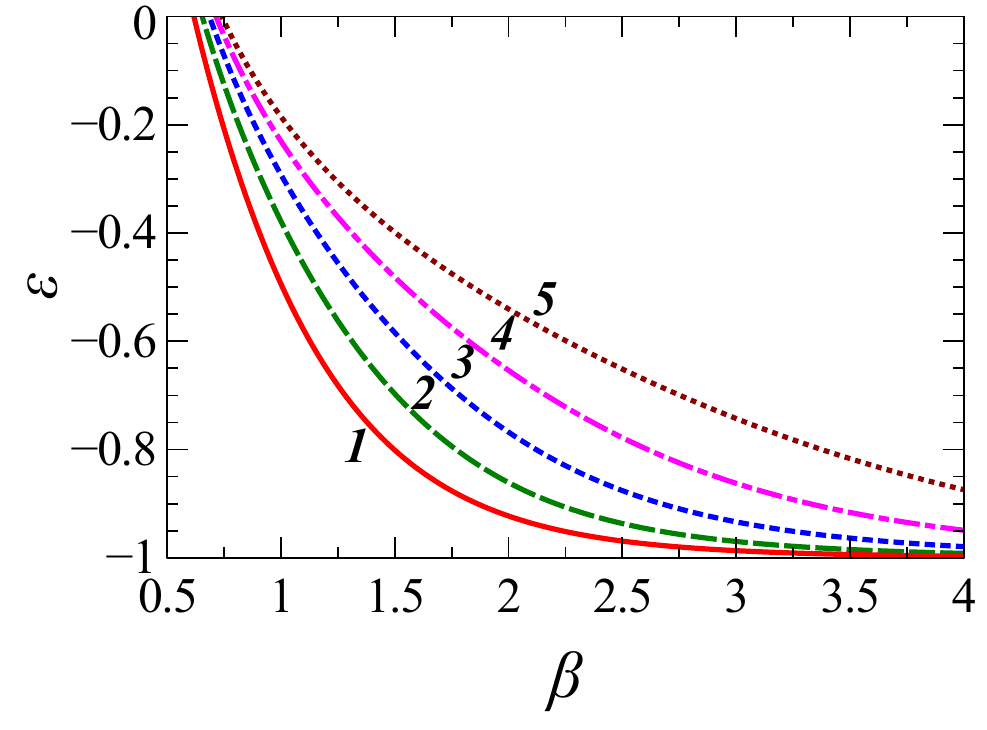}}%
%\\%
\begin{center}
\parbox[t]{0.32\textwidth}{%
\centerline{(c)}%
}%
\hspace{2cm}%
\parbox[t]{0.32\textwidth}{%
\centerline{(d)}%
}%
\end{center}
\caption{\label{fig:RPRS_obs_symm}(Color online) Fraction of good (a) and bad (b) contacts,
total adsorbed fraction (c), and energy (d) as functions of inverse
temperature $\beta$ calculated for $f_{A}=g_{\mathrm{a}}=0.5$, and various
values of cluster parameters--curve 1: $c_{\mathrm{p}}=-0.5$, $c_{\mathrm{s}}=-1$;
curve 2: $c_{\mathrm{p}}=\pm0.5$, $c_{\mathrm{s}}=\pm0.5$; curve 3: $c_{\mathrm{p}}=0$ or/and
$c_{\mathrm{s}}=0$; curve 4: $c_{\mathrm{p}}=\pm0.5$, $c_{\mathrm{s}}=\mp0.5$; curve 5: $c_{\mathrm{p}}=0.5$,
$c_{\mathrm{s}}=-1$.}
\end{figure}

Figure~\ref{fig:RPRS_obs_symm}~(a)--(c) shows the dependences of the fraction
of good (\emph{Aa}, \emph{Bb}) and bad (\emph{Ab}, \emph{Ba}) contacts
and of the total adsorbed fraction, respectively, on the inverse temperature
$\beta$. Due to the symmetry of the system, we have $\theta_{Aa}=\theta_{Bb}$
and $\theta_{Ab}=\theta_{Ba}$. With increasing $\beta$, the fraction
of good contacts grows, the maximum ``saturation'' value for both
\emph{Aa} and \emph{Bb} contacts equals 0.5, which is the maximum
that is available \emph{a priori} (according to the values of $f_{A}$
and $g_{\mathrm{a}}$), i.e., that \emph{A} (\emph{B}) monomer units and \emph{a}
(\emph{b}) surface sites may form. For the total amount of good contacts
$\theta_{\mathrm{good}}=\theta_{Aa}+\theta_{Bb}$, the upper boundary is equal
to unity. In other words, this indicates that $A\rightleftharpoons B$
and $a\rightleftharpoons b$ transformations do not occur, as opposed
to the reference system where the fraction of good \emph{A}-contacts
grows as $\beta$ increases, figure~\ref{fig:refsys}~(a). The fraction
of bad contacts, figure~\ref{fig:refsys}~(b), behaves non-monotonously:
just above the adsorption transition it grows with an increasing $\beta$
but then decays to zero. Since the fraction of bad contacts is much
lower than the fraction of good contacts, the overall adsorbed fraction
increases monotonously with $\beta$, figure~\ref{fig:RPRS_obs_symm}~(c), and the energy per monomer unit, figure~\ref{fig:RPRS_obs_symm}~(d),
monotonously decreases from 0 to $-1$.  All the curves start in the
transition points ranged in accordance with the phase diagram, figure~\ref{fig:phd}.

\subsection{Moments of monomers' and sites' distributions}

In order to understand the  behavior of the observables (which is,
as we see, much more reasonable and closer to what one can expect
in the quenched system, as compared to the reference system) better, we study
the temperature dependences of \emph{a posteriori} moments of  distributions of monomers and sites.
The first remarkable result is as follows:
in the considered symmetric case, the fractions of \emph{A} (\emph{B})
monomers and \emph{a} (\emph{b}) sites do not depend on the temperature
and are always equal to 0.5! This takes place at any combinations of correlation
parameters $c_{\mathrm{p}}$ and $c_{\mathrm{s}}$ (therefore, we do not show these
dependences in the figure due to their trivial form).

As regards the second moments of the \emph{a posteriori} distributions,
these are changing with $\beta$. This means that there are transformations
in the RC and the RS sequences but since the RC and the RS compositions
are invariant, the transformations occur according to a specific law:
\emph{A} and \emph{B} units (\emph{a} and \emph{b} sites) move in
the RC (RS) in order to tune their sequences with respect to each
other in the best way to reach the lowest interaction energy. Alternatively,
these rearrangements can be considered as coupled chemical reactions:
for example, a transformation $(\chi_{i}=A)\to(\chi_{i}=B)$ should
occur simultaneously with the reaction $(\chi_{j}=B)\to(\chi_{j}=A)$,
$i$ and $j\neq i$ denote positions of the monomer unit in the
RC sequence. At the same time, another important rule sill holds:
in the transition point, the first and the second moments of monomers
and sites distributions are equal to the corresponding \emph{a priori}
probabilities.

Now, let us consider in detail some particular cases of RC adsorption
onto RS.

\subsubsection{Quasi-blocky RC}

\begin{figure}[ht]
\includegraphics[width=4.5cm]{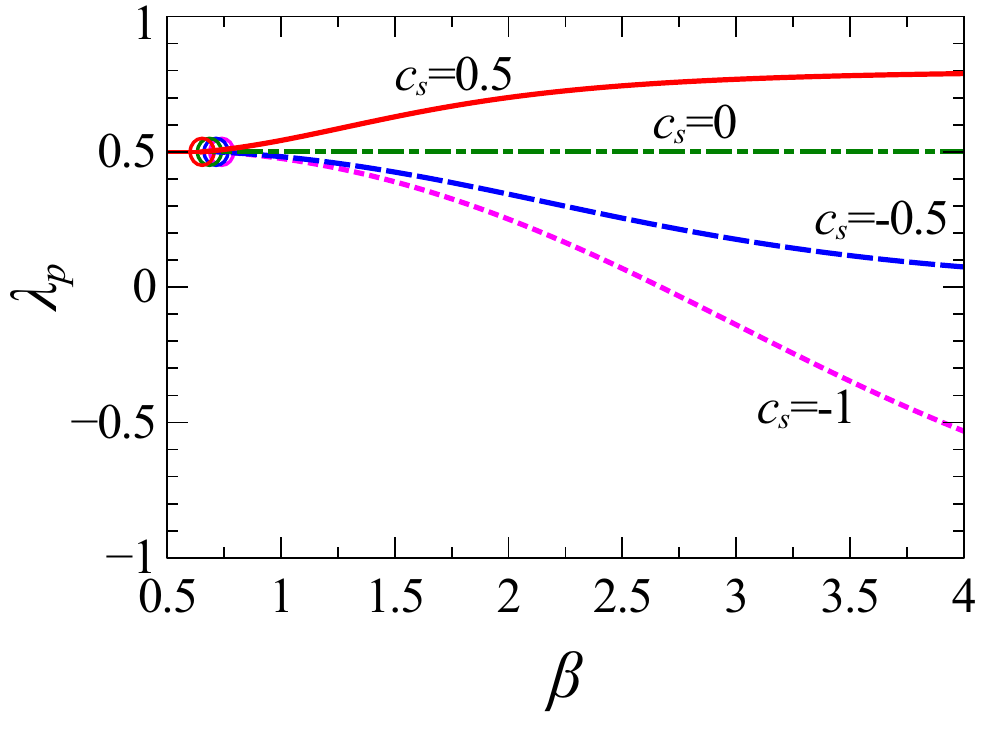}
\includegraphics[width=4.5cm]{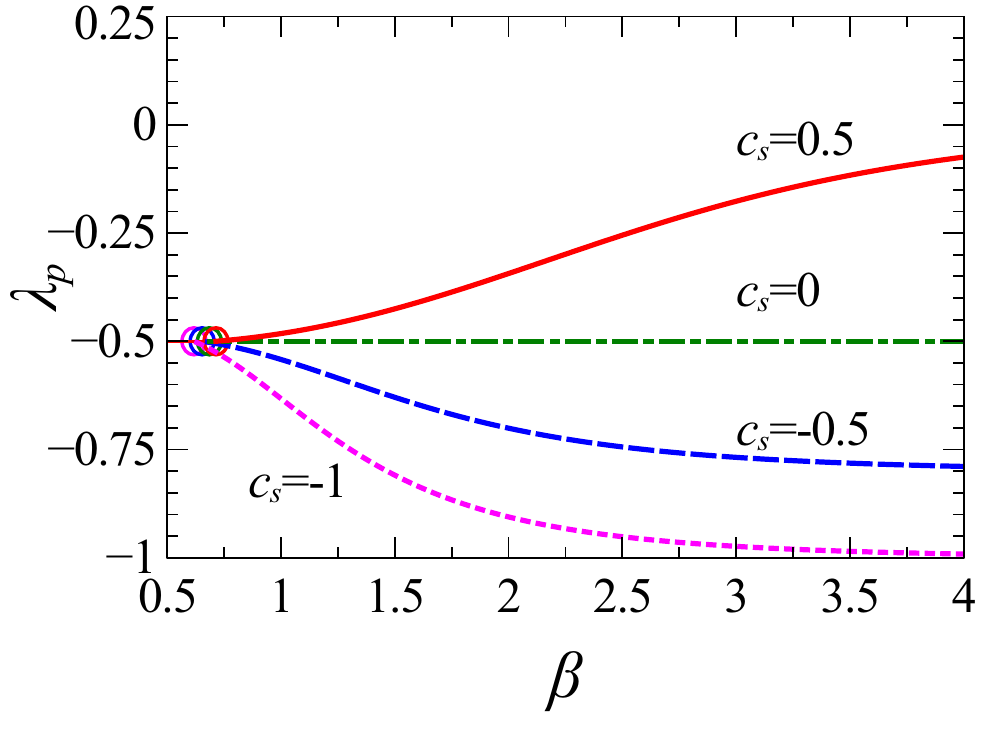}
\includegraphics[width=4.5cm]{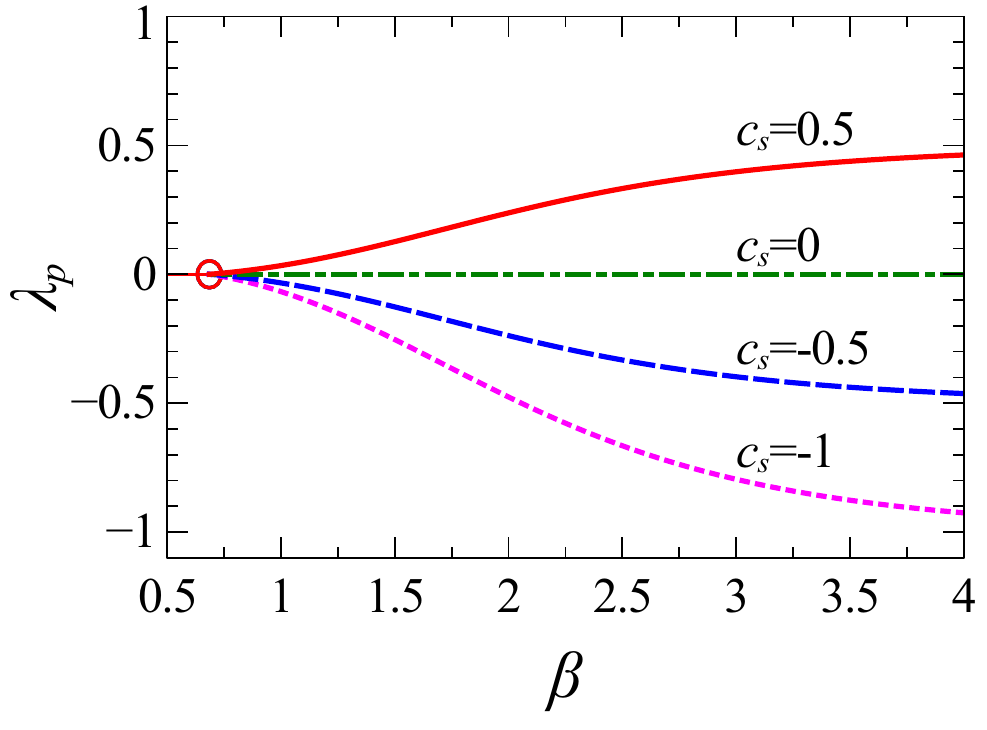}
\\%
\parbox[t]{0.32\textwidth}{%
\centerline{(a)}%
}%
\hfill%
\parbox[t]{0.32\textwidth}{%
\centerline{(b)}%
}%
\hfill
\parbox[t]{0.32\textwidth}{%
\centerline{(c)}%
}%
\hspace{5mm}
\includegraphics[width=4.5cm]{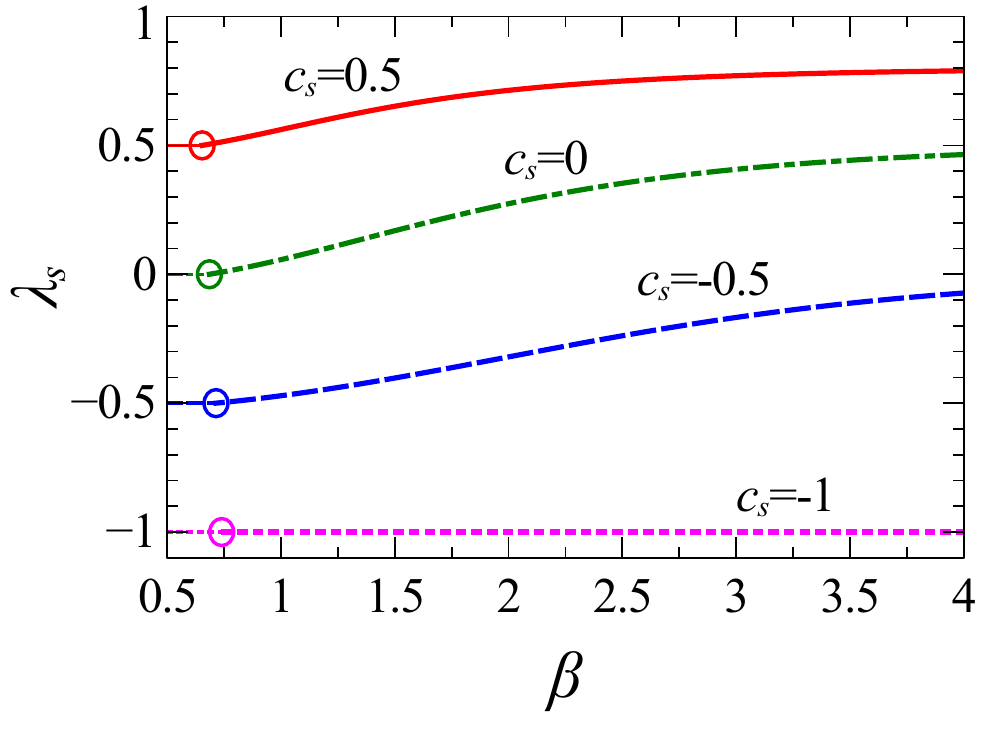}
\includegraphics[width=4.5cm]{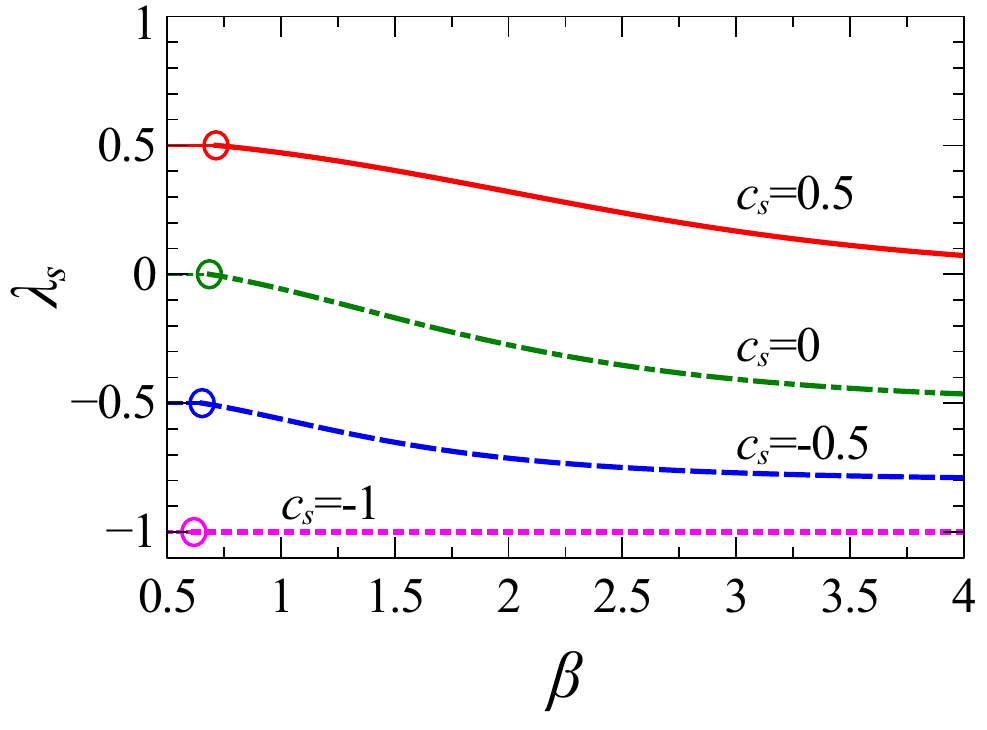}
\includegraphics[width=4.5cm]{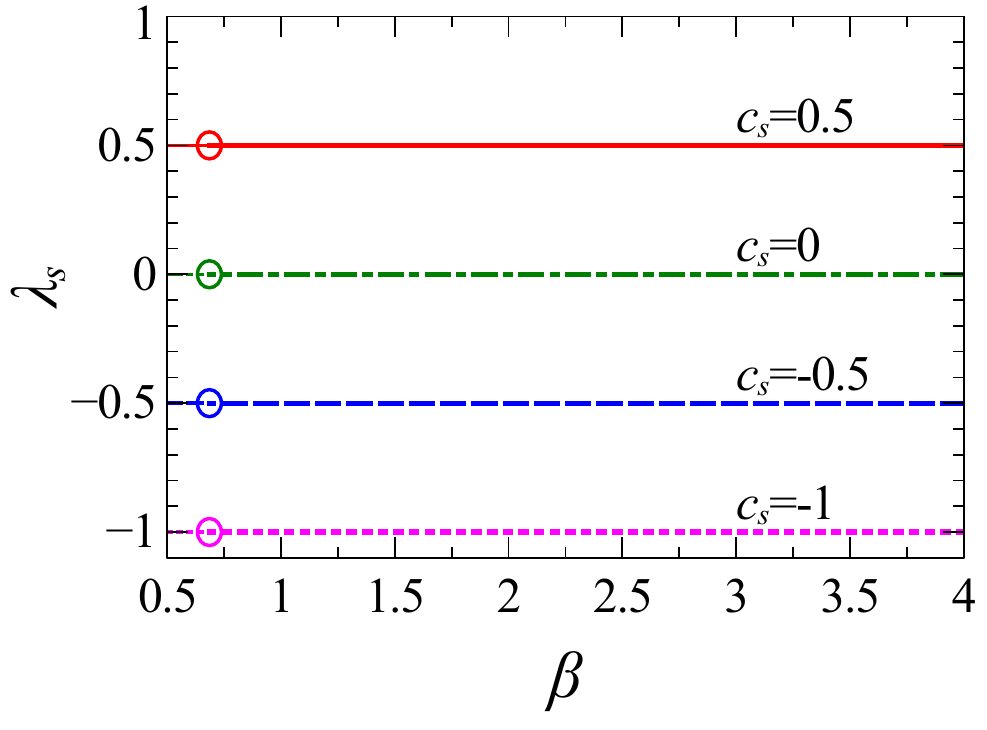}
\\%
\parbox[t]{0.32\textwidth}{%
\centerline{(d)}%
}%
\hfill%
\parbox[t]{0.32\textwidth}{%
\centerline{(e)}%
}%
\hfill
\parbox[t]{0.32\textwidth}{%
\centerline{(f)}%
}%
\hspace{5mm}
\caption{\label{fig:lambdas_qbrc}(Color online) Cluster parameters for RC (a)--(c) and RS (d)--(f)
as functions of inverse temperature $\beta$ calculated for $f_{A}=g_{\mathrm{a}}=0.5$,
$c_{\mathrm{p}}=0.5$ (a, d), $-0.5$ (b, e), 0 (c, f) and various $c_{\mathrm{s}}$ as
indicated.}
\end{figure}

As it follows from figures~\ref{fig:phd} and \ref{fig:RPRS_obs_symm},
a quasi-blocky RC has a better capability of  adsorbing onto a quasi-blocky
RS rather than onto a quasi-alternating RS. As $\beta$ increases,
the tuning of the character of correlations in both RC and RS occurs,
figure~\ref{fig:lambdas_qbrc}~(a), (d): When $c_{\mathrm{s}}=0.5\;(=c_{\mathrm{p}})$, both $\lambda_{\mathrm{p}}$ and
$\lambda_{\mathrm{s}}$ slightly increase whereas at $c_{\mathrm{s}}=-0.5\;(=-c_{\mathrm{p}})$,
both $\lambda_{\mathrm{p}}$ and $\lambda_{\mathrm{s}}$ move towards each other, i.e., the
positive $\lambda_{\mathrm{p}}$ decreases and the negative $\lambda_{\mathrm{s}}$
increases. At $c_{\mathrm{s}}=0$, an interesting behavior is observed: the
RC cluster parameter does not change but the RS cluster parameter
increases. At $c_{\mathrm{s}}=-1$, the surface regularly alternates (i.e.,
quenches) which cannot change ``by definition'', and the RC should
adjust itself; hence, $\lambda_{\mathrm{s}}$ decreases and becomes negative.

\subsubsection{Quasi-alternating RC}

Here, the same tendency to tuning as in the previous case is observed,
figure~\ref{fig:lambdas_qbrc}~(c), (f).
Hence, all the corresponding arguments may be straightforwardly reproduced
with ``the change of the sign'': at $c_{\mathrm{s}}=-0.5$ $(=c_{\mathrm{p}})$ both $\lambda_{\mathrm{p}}$ and
$\lambda_{\mathrm{s}}$ slightly decrease, at $c_{\mathrm{s}}=0.5\;(=-c_{\mathrm{p}})$ $\lambda_{\mathrm{p}}$
and $\lambda_{\mathrm{s}}$ move towards each other, at $c_{\mathrm{s}}=0$ $\lambda_{\mathrm{p}}$
does not change and $\lambda_{\mathrm{s}}$ decreases, at $c_{\mathrm{s}}=-1,$ $\lambda_{\mathrm{p}}$
decreases towards $-1$.

\subsubsection{Bernoullian RC}

With respect to the transition point and the adsorbed amount, figures~\ref{fig:phd} and \ref{fig:RPRS_obs_symm}, the behavior of the annealed
Bernoullian copolymer does not depend on the correlations in the annealed
RS (the latter can be even quenched regularly alternating with $c_{\mathrm{s}}=-1$).
For the second moments of monomers and sites distributions in the
RC and RS, respectively, the following regularity  is observed: the\emph{
a posteriori} cluster parameters $\lambda_{\mathrm{s}}$ in the RS --- the ``partner''
of the Bernoullian RC --- does not change with $\beta$ and remains equal
to the \emph{a priori} correlation parameter $c_{\mathrm{s}}$,
figure~\ref{fig:lambdas_qbrc}~(b), (e). The ``double
Bernoullian'' case $c_{\mathrm{p}}=c_{\mathrm{s}}=0$ is remarkable: the second moment
does not change both in RC and RS.

\subsubsection{Summary for the symmetric case}

The analysis of the first and the second moments of monomers and site distributions
in RC and RS, respectively, in the symmetric case shows that upon
a decrease in temperature (increase in $\beta$), the increase in
the number of good contacts and the corresponding decrease in the
number of bad contacts is implemented via rearrangement of \emph{A}
and \emph{B} monomer units (\emph{a} and \emph{b} sites) in the RC
chain (on the RS). We have also seen that the first moments (and,
in some cases, the second moments) keep their expected values and,
in this sense, the annealed approximation turns out to be more accurate
than in the case of the (more simple) reference system. This also
means that an improvement of the annealed approximation with the aid
of the first-order Morita approximation (and the second order--in
the case of the Bernoullian RC and RS) will produce no effect because
the corresponding Morita constraints are already satisfied in the
annealed system and the same results will be obtained.

\subsection{The effect of asymmetry}

As we have seen, in the symmetric case, the  adsorption of annealed RC onto
annealed RS exhibits the most interesting features, the most essential
among them being the correct (with respect to the \emph{a priori} probabilities)
values of the first moments of monomers and the distributions of sites  meaning
that the overall composition of the RC and RS does not change during
the interaction. This is the consequence of the system symmetry. When
both the RC and the RS are, in addition, Bernoullian, the second moments
keep their correct values too.

Let us now analyze how a violation of the system's symmetry affects
the RC adsorption onto the RS. In principle, the symmetry can be destroyed
in two different ways: one can either (1) choose non-equal probabilities
for \emph{A} and \emph{B} units (\emph{a} and \emph{b} sites) to appear
in the RC sequence (on the RS), i.e., set $f_{A}$, $g_{\mathrm{a}}\neq0.5$
or (2) change the value of one of the interaction parameters $\epsilon_{ij}$.
To see how different ways of introducing the asymmetry influence the
adsorption, let us take the (simplest) set of parameters with $f_{A}=g_{\mathrm{a}}=0.5$,
$c_{\mathrm{p}}=c_{\mathrm{s}}=0$ (both RC and RS are Bernoullian), $\epsilon_{Aa}=\epsilon_{Bb}=-1$,
and $\epsilon_{Ab}=\epsilon_{Ba}=0$ as the reference. We will change
one of the composition or the interaction parameters, keeping other parameters unchanged,
as in the reference system. In particular, we (1) slightly change
the probabilities of the appearance of \emph{A} and \emph{B} monomers
in the RC, $f_{A}=0.6$, keeping for the RS $g_{\mathrm{a}}=0.5$ or (2) increase
the attraction of \emph{A} monomer to a site by setting $\epsilon_{Aa}=-1.2$
or (3) make one of the bad contacts repulsive instead of neutral by
setting $\epsilon_{Ab}=+1$, or (4) turn the good \emph{Bb} contacts
into bad by setting $\epsilon_{Bb}=\epsilon_{Ab}=\epsilon_{Ba}=0$.

\begin{figure}[ht]
\includegraphics[width=4.5cm]{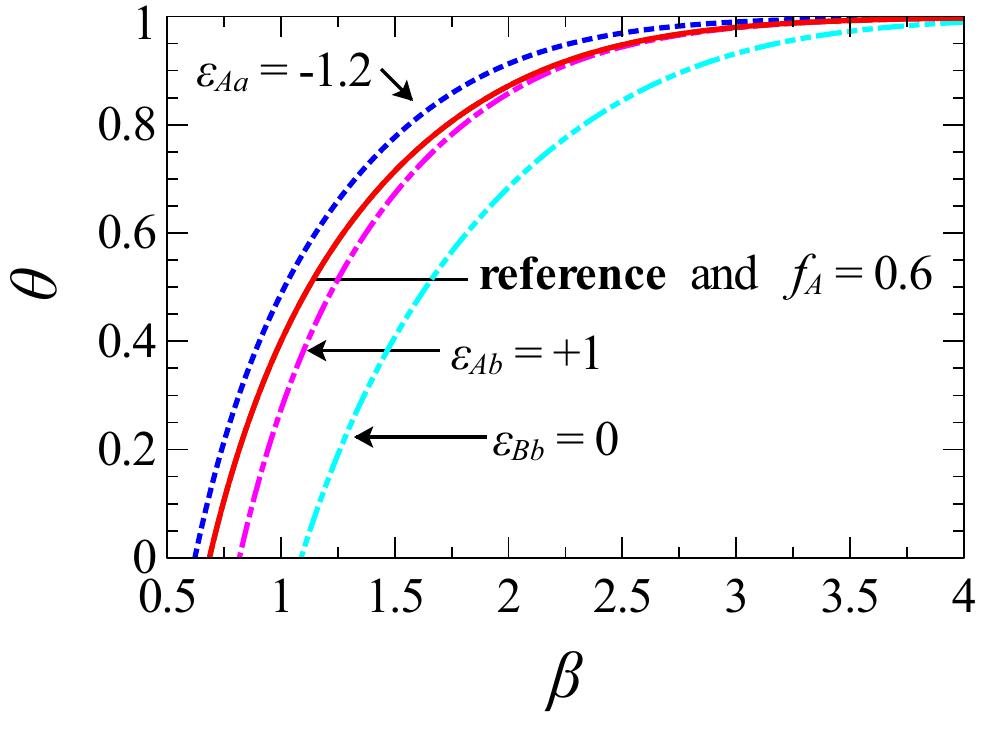}
\includegraphics[width=4.5cm]{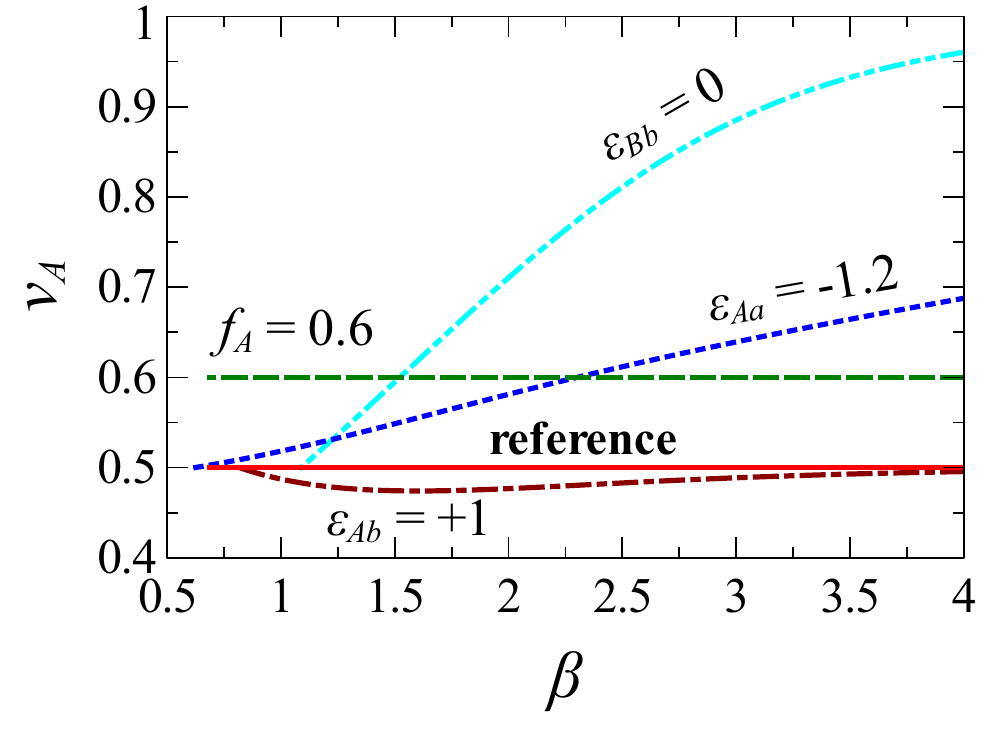}
\includegraphics[width=4.5cm]{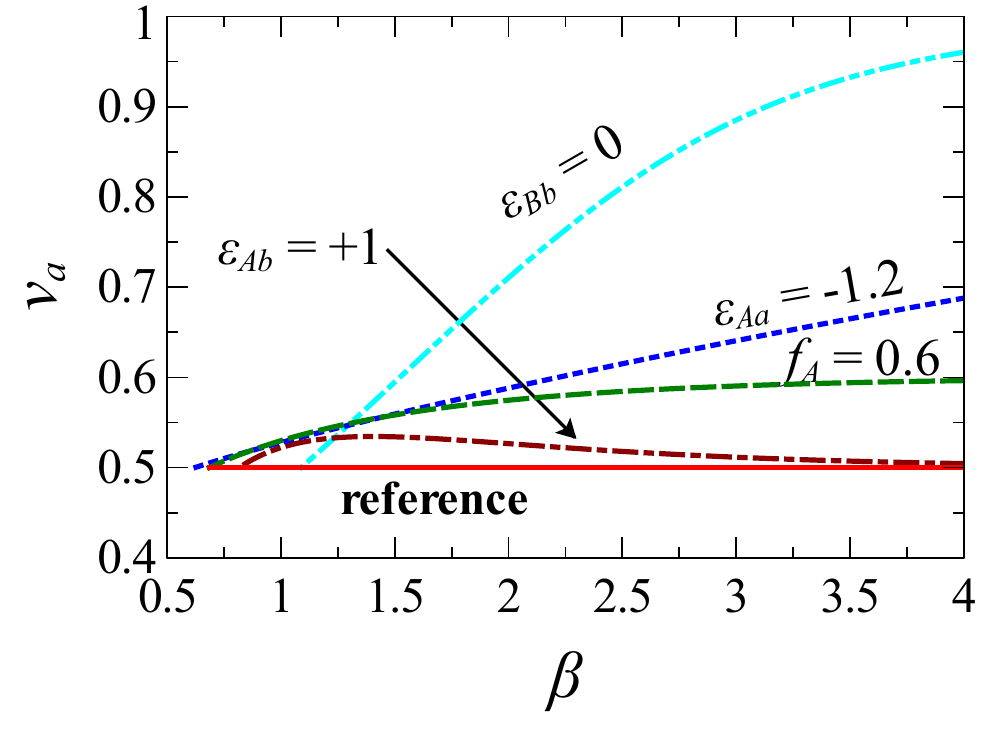}
\\%
\parbox[t]{0.32\textwidth}{%
\centerline{(a)}%
}%
\hfill%
\parbox[t]{0.32\textwidth}{%
\centerline{(b)}%
}%
\hfill
\parbox[t]{0.32\textwidth}{%
\centerline{(c)}%
}%
\hspace{5mm}\caption{\label{fig:asymmetry-1}(Color online) Overall adsorbed fraction (a), fraction of
\emph{A}-monomers (b) and \emph{a}-sites (b) as functions of inverse
temperature $\beta$ calculated for symmetric reference system (red
solid curves) with $f_{A}=g_{\mathrm{a}}=0.5$, $c_{\mathrm{p}}=c_{\mathrm{s}}=0$, $\epsilon_{Aa}=\epsilon_{Bb}=-1$,
and $\epsilon_{Ab}=\epsilon_{Ba}=0$. Other curves are for the systems
differing from the reference one by one of the parameters, indicated
at the curve.}
\end{figure}
These four cases are compared in figure~\ref{fig:asymmetry-1}. The
results of the comparison are as follows: (1) an increase of the probability
of \emph{A} monomers present in the sequence, changes neither the position
of the adsorption transition point nor the shape of the temperature
dependence of the adsorbed fraction (and of the GF smallest singularity).
At the same time, the RC and
the RS compositions behave differently with the change of temperature:
as the inverse temperature $\beta$ increases, composition of the
RC remains the same whereas the fraction of \emph{a}-sites increases
and tends to 0.6. (2) An increase in the \emph{Aa} affinity results
in the shift of the transition point toward smaller $\beta$, while with
an increase of $\beta$, the (expected) growth is observed in the amount of \emph{A}
monomers and \emph{a} sites providing the most favorable \emph{Aa}
and \emph{Bb} contacts. (3) An increase in the \emph{Ab}
interaction energy, shifts the transition point to larger values of
$\beta$; fractions of \emph{A} monomers and \emph{B} sites
slightly decrease just after the adsorption transition; as $\beta$ increases,
they both  tend to 0.5, in order to have a maximum of possible favorable
\emph{Aa} and \emph{Bb} contacts. (4) The transformation of attractive
\emph{Bb} contacts intro neutral ones shifts the transition point
towards larger $\beta$ and leads to massive $B\to A$ and $b\to a$
transformations with an increasing $\beta$.

The observed behavior is in agreement with the behavior of the reference
system (RC adsorbing onto a homogeneous surface, see section~\ref{sec:refsys}),
where the equilibrium is shifted towards transformation of non-adsorbing
monomer units into adsorbing ones. In our case, we encounter more complicated
coherent monomers and transformations of sites. As soon as some preferences
appear in the transformation reaction constants (governed by $f_{A}$
and $g_{\mathrm{a}}$) or in the interaction map, the equilibrium shifts toward
these preferences.

\begin{figure}[t]
\centerline{
\includegraphics[width=0.32\textwidth]{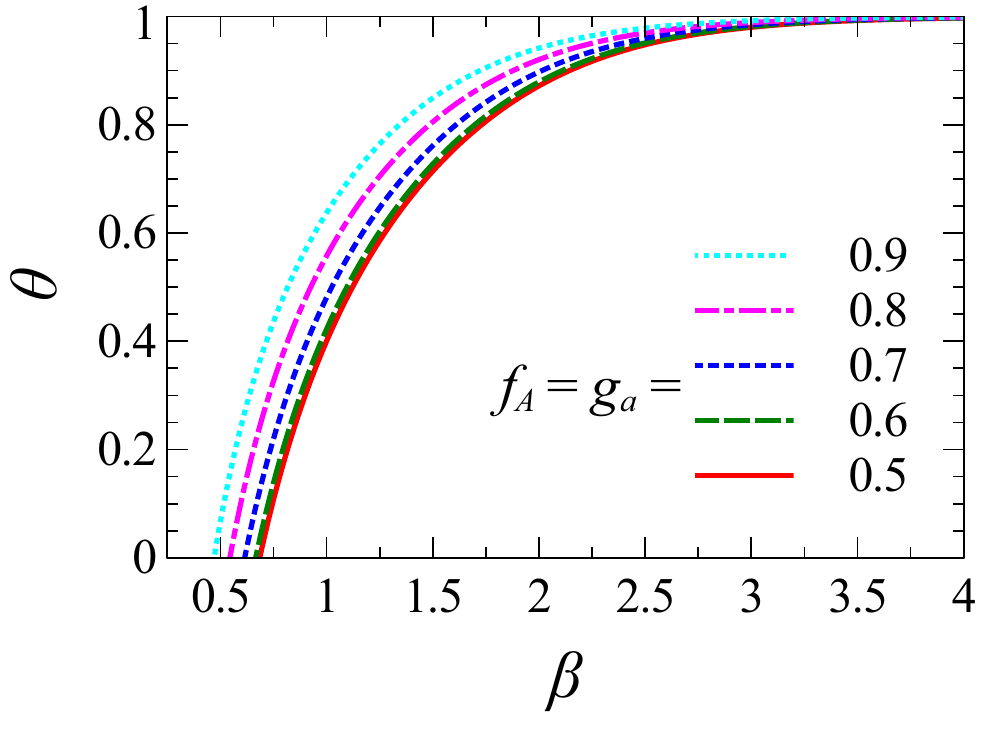}%
\hspace{2cm}%
\includegraphics[width=0.32\textwidth]{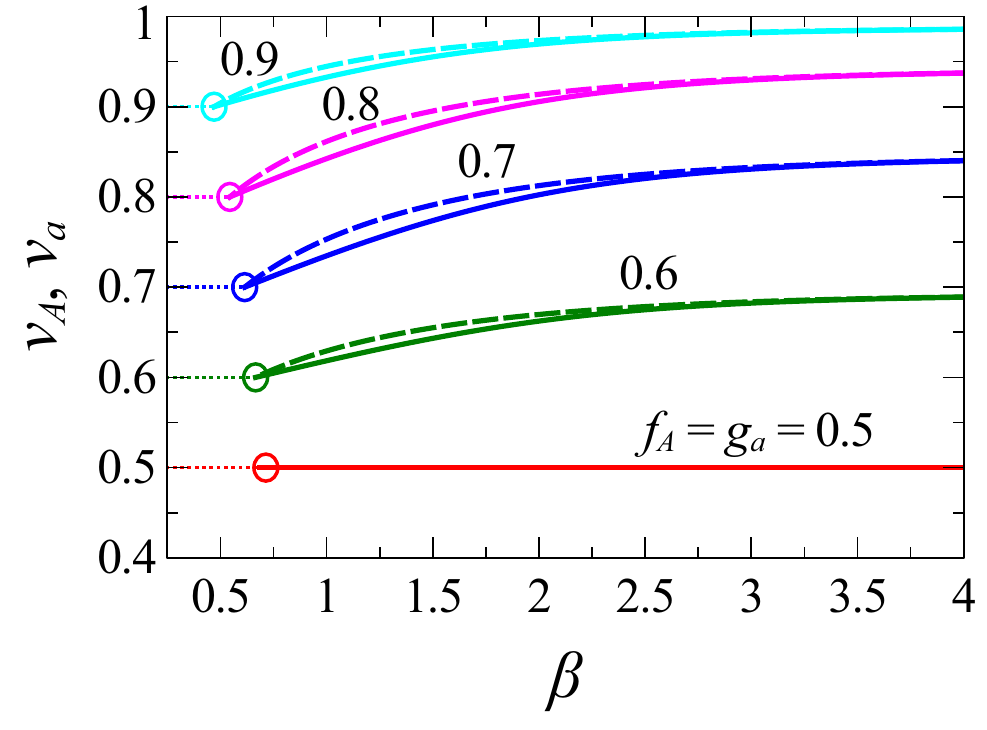}}%
%\\%
\begin{center}
\parbox[t]{0.32\textwidth}{%
\centerline{(a)}%
}%
\hspace{2cm}%
\parbox[t]{0.32\textwidth}{%
\centerline{(b)}%
}%
\end{center}
\caption{\label{fig:asymmetry-2}(Color online) Overall adsorbed fraction (a) and fractions
of \emph{A}-monomers (solid curves) and \emph{a}-sites (dashed curves)
(b) as functions of inverse temperature $\beta$ calculated for $c_{\mathrm{p}}=c_{\mathrm{s}}=0$,
$\epsilon_{Aa}=\epsilon_{Bb}=-1$, $\epsilon_{Ab}=\epsilon_{Ba}=0$,
and equal fractions of \emph{A}-monomers and \emph{a}-sites ($f_{A}=g_{\mathrm{a}}$)
as indicated.}
\end{figure}

There is another interesting way to partly violate the symmetry of
the system. By comparing the fraction of \emph{A}-monomers with that
of \emph{a}-sites and the fraction of \emph{B}-monomers with that
of \emph{b}-sites we can obtain the upper boundary for the maximum
possible fraction of good contacts. Since good contacts are the
\emph{Aa} and the \emph{Bb} ones, then $\theta_{\mathrm{good}}=\theta_{Aa}+\theta_{Bb}$.
Obviously, $\theta_{Aa}\leqslant \min(f_{A},g_{\mathrm{a}})$ , $\theta_{Bb}\leqslant \min(f_{B},g_{b})$,
hence, $\theta_{\mathrm{good}}^{\mathrm{max}}=\min(f_{A},g_{\mathrm{a}})+\min(f_{B},g_{b})$.
In the symmetric case, $f_{A}=f_{B}=g_{\mathrm{a}}=g_{b}=0.5$, therefore, $\theta_{\mathrm{good}}^{\mathrm{max}}=1$.
If we choose the RC and the RS compositions, so that $f_{A}=g_{\mathrm{a}}$
and $f_{B}=g_{b}$, the upper boundary for the total fraction of good
contacts will still be equal to one, as it is in the symmetric case.

Temperature dependences of the overall adsorbed fraction, RC and RS
compositions and cluster parameters for $f_{A}=g_{\mathrm{a}}\geqslant 0.5$ and
$c_{\mathrm{p}}=c_{\mathrm{s}}=0$ are presented in figure~\ref{fig:asymmetry-2}. We
see that in spite of the invariance of $\theta_{\mathrm{good}}^{\mathrm{max}}$, a simultaneous
increase in the fraction of \emph{A}-monomers and \emph{a}-sites favors
adsorption: the transition point shifts to the lower $\beta$ values,
for a given temperature, the total adsorbed fraction is larger in
the case of larger $f_{A}=g_{\mathrm{a}}$. Moreover, with an increasing $\beta$,
the fractions of the major \emph{A} and \emph{a} components grow and
the fractions of the minor \emph{B} and \emph{b} components correspondingly
decrease. It is also clear that if we take $f_{A}=g_{\mathrm{a}}<0.5$, the
picture will remain qualitatively and quantitatively the same with
respect to major and minor components (here, \emph{B} and \emph{b}
are major components, \emph{A} and \emph{a} are minor components). The only exception
is the symmetric case $f_{A}=g_{\mathrm{a}}=0.5$ , where there is a strict
\emph{A/B} and \emph{a/b} balance. That is, one can say that the equilibrium
gets shifted towards the major component, and the situation becomes
similar in some sense to that in the reference system of section~\ref{sec:refsys}.

\section{\label{sec:summary} Conclusion}

We have considered a two-dimensional partially directed walk (2D-PDW)
model of a two-letter (\emph{AB}) random copolymer (RC) adsorption
onto a two-letter (\emph{ab}) random surface (RS). This model was
introduced in our previous work \cite{Polotsky:2012} where it was
treated by using the combination of the annealed approximation
(to perform double averaging over the sequence and surface disorder) with
the generating functions (GFs) approach (to  sum over polymer conformations).
In contrast to \cite{Polotsky:2012}, in the present work we have
gone beyond the transition point and studied the temperature dependences
of various observables for the annealed symmetric system. This choice
was motivated for several reasons: The annealed approximation provides a zero-order
rough approximation to a realistic quenched system; on the other hand,
it can serve as a prototype of real physical systems like two-state
polymers \cite{Yoshinaga:2008}. The system was chosen to be ``two-fold''
symmetric: with respect to the composition of the RC and RS and with respect
to the interaction map (good attractive \emph{Aa} and \emph{Bb} contacts
had the same energy, while both bad \emph{Ab }and \emph{Ba} contacts
were equally neutral). In \cite{Polotsky:2012} it was shown that
in this case the system has the most interesting phase diagram. Finally,
the symmetry of the system makes highly unpredictable interconversion of monomers and sites ($A\rightleftharpoons B$ and $a\rightleftharpoons b$)
in the annealed system. Therefore, special attention was paid to \emph{a
posteriori} moments of  distributions of monomers and sites.

We have shown that in the considered symmetric case, the expected \emph{a
posteriori} compositions of the RC and the RS correspond to the\emph{
a priori} probabilities to meet \emph{A} monomers and a sites in the
RC monomer sequence / RS site sequence and do not change with the
temperature. At the same time, the \emph{a posteriori} cluster parameter
(related to the second moments of  distributions of monomers and sites)
in the RC and RS changes with the temperature, indicating that monomers
and sites rearrange in the RC and the RS to provide a better matching
between them and, hence, a stronger adsorption. A special case is
the one where both the RC and the RS are Bernoullian: in this situation,
both first and second moments keep their correct values at any temperature.

We have also studied the effect of the system symmetry violation on
the adsorption behavior. There are various ways of doing this and
all of them shift the equilibrium towards the major component and/or more
favorable contacts.

\section*{Acknowledgements}

The author is grateful to Dr. Oleg V. Borisov for his critical comments
and suggestions. This work was partially supported by the Government
of the Russian Federation, grant 074--U01.

%% Type in your references using {thebibliography} environment
%% or create them from your bibtex database using cmpj.bst style (experimental).

% \bibliographystyle{cmpj}
% \bibliography{dpa}

%
%% If you have problems with typesetting in ukrainian uncomment lines below.
%
%  \lastpage
%  \end{document}

\ukrainianpart

\title{Адсорбція симетричного випадкового кополімера на симетричну випадкову поверхню: відпалений випадок}
\author{A.A. Полоцький\refaddr{label1,label2}}

\addresses{
\addr{label1} Інститут макромолекулярних сполук, Російська академія наук,
  199004 Санкт-Петербург, Російська Федерація
\addr{label2} Санкт-Петербурзький національний дослідницький університет інформаційних технологій, механіки та оптики (університет ITMO),
  197101 Санкт-Петербург, Російська Федерація
}

\makeukrtitle

\begin{abstract}
\tolerance=3000%
Адсорбція симетричного (\emph{AB}) випадкового кополімера (ВК) на симетричну (\emph{ab}) випадкову неоднорідну поверхню (ВП) вивчається
у наближенні відпалу із використаннями для полімера двовимірної моделі частково напрямлених блукань.
Показано, що у симетричному випадку очікувані \emph{a posteriori} концентрації ВК і ВП мають правильні значення  (які відповідають  їх \emph{ a priori} ймовірностям) і не змінюються, в залежності від температури, в той час, як другі моменти розподілів мономерів і вузлів в ВК і ВП змінюються.
Це показує, що мономери і вузли взаємно не перетворюються, а лише перегруповуються, щоб забезпечити краще допасовування  між ними і, як результат, сильнішу адсорбцію  ВК на ВП. Проте будь-яке порушення симетрії системи зсуває рівновагу у напрямку основної компоненти і/або до більш сприятливих контактів та приводить  до взаємоперетворення мономерів і вузлів.

\keywords випадковий кополімер, випадкова поверхня, полімерна адсорбція, наближення відпалу, генеруючі функції

\end{abstract}

\end{document}